\newif\ifShowKeys
\ShowKeysfalse

\documentclass[11pt,a4paper]{article} 
\pdfoutput=1

\usepackage{booktabs} 
\usepackage{colortbl} 
\usepackage{xcolor}

\usepackage[no-natbib-sort]{my-jheppub}

\usepackage{amsmath, amssymb}

\usepackage{mathpazo}
\usepackage{bm}
\usepackage{environ}
\usepackage{mathrsfs}
\usepackage{array,arydshln}

\usepackage{graphicx,epsfig}
\usepackage{tikz}							
\usepackage{epic}
\usepackage{youngtab}
\usepackage{float}
\usepackage{color}
\definecolor{maroon}{rgb}{0.8,0.3,0.}

\usepackage{slashed}
\usepackage[nodayofweek]{date time}

\ifShowKeys \usepackage[notcite]{showkeys} \fi

\usepackage{hyperref}

\usepackage{aurical}
\usepackage[T1]{fontenc}

\usepackage{framed}
\definecolor{shadecolor}{RGB}{255, 230, 204}

\allowdisplaybreaks

\newcommand{\be}{\begin{equation}}
\newcommand{\ee}{\end{equation}}

\newcommand{\mc}{\mathcal }

\newcommand{\bv}{\begin{verbatim}}
\newcommand{\ev}{\end{verbatim}}

\newcommand{\la}{\label}

\newcommand{\eps}{\varepsilon}
\newcommand{\wt}{\widetilde}
\newcommand{\E}{\text{E}}
\newcommand{\inst}{^{\text{inst}}}
\newcommand{\pert}{^{\text{pert}}}

\def\confblock(#1,#2,#3,#4,#5,#6,#7) { #1_{#2}\!
{\small\begin{bmatrix} #4 & #5 \\ #3 & #6  \end{bmatrix}}\!(#7) }

\title{Exact partition functions for deformed $\mc N=2$ theories with $N_{f}=4$ flavours}
\author[a,b]{Matteo Beccaria,} 
\author[a,b]{Alberto Fachechi,} 
\author[a,b]{Guido Macorini,} 
\author[a,b]{and Luigi Martina} 

\abstract{
We consider the $\Omega$-deformed $\mathcal{N}=2$ $SU(2)$ gauge theory in four dimensions
with $N_{f}=4$ massive fundamental hypermultiplets. The low energy effective action depends on 
the deformation parameters $\eps_{1}, \eps_{2}$, the scalar field expectation value $a$, and the hypermultiplet masses $\bm{m}=(m_{1}, m_{2}, m_{3}, m_{4})$. Motivated by recent findings in the $\mathcal{N}=2^{*}$
theory, we explore the theories that are characterized by special fixed 
ratios $\eps_{2}/\eps_{1}$ and $\bm{m}/\eps_{1}$ and propose a simple condition 
on the structure of the multi-instanton contributions to the prepotential determining the 
effective action.
This condition determines  a finite  set $\Pi_{N}$ of special points such that 
the prepotential has $N$ poles at fixed positions independent on the instanton number. In 
analogy with what happens in the 
$\mathcal{N}=2^{*}$ gauge theory, the full prepotential of the $\Pi_{N}$ theories may be given in closed form as an explicit function of $a$ and the modular parameter $q$ appearing in special combinations of 
 Eisenstein series and Jacobi theta  functions with well defined
modular properties.  The resulting finite pole partition functions are related by AGT correspondence
to special 4-point spherical conformal blocks of the Virasoro algebra. We examine in full details 
special cases where the closed expression of the block is known and confirms our 
Ansatz. We systematically study the special features of Zamolodchikov's recursion for the $\Pi_{N}$ conformal blocks. As a result, 
we provide a novel effective recursion relation that can be exactly solved and allows to prove
 the conjectured closed expressions analytically in the case of the $\Pi_{1}$ and $\Pi_{2}$ 
 conformal blocks. 
\vfill }

\affiliation[a]{Dipartimento di Matematica e Fisica Ennio De Giorgi,\\
Universit\`a del Salento, Via Arnesano, 73100 Lecce, 
Italy} 

\affiliation[b]{INFN, Via Arnesano, 73100 Lecce, Italy}

\emailAdd{matteo.beccaria@le.infn.it} 

\begin{document}



\maketitle
\flushbottom


\section{Introduction and results}

In this paper, we consider the four dimensional $\mc N = 2$ $SU(2)$ theory with $N_{f}$ 
flavour hypermultiplets in the fundamental representation of the gauge group. 
At the special value $N_{f}=4$ the one loop $\beta$-function 
vanishes and  conformal invariance is broken  by the flavour masses 
$\bm{m}=(m_{1}, \dots, m_{4})$.
Due to $\mc N=2$ supersymmetry, 
the full effective action may be expressed in terms of the analytic prepotential $\mc F(a, \bm{m})$,
where $a$ is the  vacuum expectation value  of the scalar 
in the adjoint gauge multiplet \cite{D'Hoker:1999ft}. 
Non-perturbative effects are present due to instantons and they are captured by the exact 
Seiberg-Witten (SW) solution \cite{Seiberg:1994rs,Seiberg:1994aj}. The instanton expansion 
is weighted by the instanton counting parameter $x$ which is a cross-ratio of the four roots of the original SW curve. In the following, it will be convenient to trade $x$ by the modular parameter $q$ defined by, see App.~(\ref{sec:ids}),
\be
\label{1.1}
q = \exp\bigg(-\pi\,\frac{\mathbb K(1-x)}{\mathbb K(x)}\bigg),
\ee
where $\mathbb K$ is the complete elliptic integral of the first kind.
Indeed, the quantity $\tau$ appearing in the parametrization $q=e^{i\,\pi\,\tau}$ receives only corrections proportional to the hypermultiplet masses.
It is the analogue of the complexified gauge coupling constant of the related $\mc N=2^{*}$ theory
whose matter content is a massive adjoint hypermultiplet (see  \cite{Billo:2010mg}
for a detailed  recent discussion).

\medskip
As it is well known, an important alternative to the SW approach is localization 
\cite{Nekrasov:2002qd,Nekrasov:2003rj,Nakajima:2003uh}. 
The many-instanton moduli space is regularized by considering the theory on 
the $\Omega$-background,
{\em i.e.} a 4d Poincar\' e breaking deformation 
depending on two parameters ${\bm{\eps}=(\eps_{1}, \eps_{2})}$ 
\cite{Flume:2002az,Bruzzo:2002xf,Flume:2004rp,Nekrasov:2004vw,Marino:2004cn,
Billo:2009di,Fucito:2009rs,Billo:2010bd}. \footnote{
For the string interpretation of the $\Omega$-background and its BPS excitations
see \cite{Hellerman:2011mv,Hellerman:2012zf,Hellerman:2012rd,Orlando:2013yea}.
The approach in these papers provides a geometric interpretation for the localization 
(in terms of a dilaton potential), and gives a clear understanding of all the possible deformation parameters 
($\eps$ and mass terms). It also gives a way to write down an effective action that contains 
exactly the same information as the partition function.
}
In this approach, the deformed 
partition function $Z\inst(\bm{\eps}, a, \bm{m})$ is well defined and its associated 
non-perturbative $\eps$-deformed prepotential is introduced according to  
\be
\la{1.2}
F\inst(\bm{\eps}, a, \bm{m})= -\eps_1\eps_2\,\log Z\inst(\bm{\eps}, a, \bm{m}).
\ee
Besides, the complete prepotential include a (simpler) perturbative contribution 
$F\pert$ 
taking into account the 1-loop renormalization of the gauge coupling by terms proportional 
to the masses $\bm{m}$. Due to superconformal invariance, this is the full perturbative effect.
The partition function in the $\eps$-background is an interesting object because its expansion 
around vanishing 
deformation parameters generates higher genus amplitudes of the $\mc N=2$
topological string  \cite{Antoniadis:1993ze,Antoniadis:2010iq,Krefl:2010fm,
Huang:2010kf,Antoniadis:2013mna,Antoniadis:2013epe,Florakis:2015ied} and satisfies a 
holomorphic anomaly equation \cite{Bershadsky:1993ta,Bershadsky:1993cx,Klemm:2002pa,Huang:2009md}.

\medskip
At finite values of $\eps_{1}, \eps_{2}$, the deformed partition function is a central quantity 
in  the Alday-Gaiotto-Tachikawa (AGT) correspondence 
\cite{Alday:2009aq}. \footnote{
A simpler setup is the Nekrasov-Shatashvili (NS) limit \cite{Nekrasov:2009rc} 
where one of the two $\eps$ parameters
vanishes. The supersymmetric vacua of the deformed  theory 
are related to the  eigenstates of a quantum integrable system. In the  case of the 
$\mc N=2^{*}$ theory, the relevant 
integrable system is the elliptic Calogero-Moser system \cite{Nekrasov:2009rc} and the associated
spectral problem reduces to the study of the celebrated Lam\'e equation.
If the  hypermultiplet mass $m$ is taken to be proportional to $\eps$ with definite special ratios
$\tfrac{m}{\eps} = n+\tfrac{1}{2}$, where $n\in\mathbb{N}$, the spectral problem is $n$-gap
and major simplifications
occur in the $k$-instanton prepotential contributions  \cite{Beccaria:2016wop}.
}
This is a map between deformed $\mc N=2$  instanton partition functions  
and conformal blocks of a suitable CFT with certain  worldsheet genus and 
operator insertions. AGT correspondence may be checked perturbatively in the 
number of instantons \cite{Poghossian:2009mk,Fateev:2009aw,Alba:2010qc} and it has been partially proved
in \cite{Mironov:2009ib,Mironov:2010zs,Mironov:2010su,Mironov:2010xs,Mironov:2010pi,Morozov:2013rma}, see also \cite{Maruyoshi:2014eja}.
For the 
$\mc N=2^{*}$ $\eps$-deformed $SU(2)$ gauge theory 
the relevant CFT quantity is the one-point conformal block on the 
torus.
In the case of the $SU(2)$ theory with $N_{f} = 4$ the generalized prepotential is  related to the 
logarithm of the conformal blocks of four Liouville operators on a sphere with the bare gauge coupling 
constant being associated with the cross-ratio of the punctures where the four operators are inserted
\cite{Fateev:2009me,Poghossian:2009mk,Hadasz:2009db,Fateev:2009aw,Menotti:2010en,Marshakov:2010fx,KashaniPoor:2012wb,Piatek:2013ifa,Kashani-Poor:2014mua,Alkalaev:2016ptm}.

\medskip
The AGT perspective suggests to inspect the  modular properties that have been studied 
in the undeformed case  \cite{Minahan:1997if,Billo:2011pr} as well as for $\eps_{1,2}\neq 0$
\cite{Huang:2011qx,Huang:2012kn}. It is possible to sum in closed form 
the instanton expansion order by order at large vacuum expectation value $a\to \infty$. In more details, 
for the $\mc N=2^{*}$ theory, 
the coefficients of the various $(1/a)^{n}$ terms are written in  terms of 
quasi-modular Eisenstein series depending on $q$. Similar results are available for the $N_{f}=4$ theory
by adding contributions involving the Jacobi theta functions $\vartheta_{2}$, $\vartheta_{4}$.
This approach has been successful in the gauge theory  
\cite{Billo:2011pr,Billo:2013fi,Billo:2013jba,Billo:2014bja,Billo:2015ria,Billo:2015jta,Billo:2016zbf,Ashok:2016oyh}, in CFT language by AGT correspondence
 \cite{KashaniPoor:2012wb,Piatek:2013ifa,Kashani-Poor:2013oza,Kashani-Poor:2014mua}, and by 
a semiclassical WKB analysis
 \cite{Mironov:2009dv,Mironov:2009uv,He:2010xa,He:2010if,Popolitov:2013ria,He:2014yka}.

\medskip
In a recent paper \cite{Beccaria:2016nnb}, the $\Omega$-deformed 
partition function of the $\mc N=2^{*}$ theory is analyzed 
in the $(\alpha, \beta)$ plane where $\alpha, \beta$ are 
the ratios $\alpha=m/\eps_{1}$ ($m$ being the hypermultiplet mass) and $\beta=\eps_{2}/\eps_{1}$.
At each instanton number, the partition function is a rational function of the 
ratio $\nu=2\,a/\eps_{1}$. Special points $(\alpha, \beta)$ do exist with the property that 
the $k$-instanton prepotential has poles at a fixed set of 
positions $\nu\in\{\nu_{1}, \dots, \nu_{N}\}$ independent on $k$.
At these special $N$-poles points,  the instanton
partition function and the perturbative part of the prepotential take an exact closed 
form to be reviewed in Sec.~(\ref{sec:review}).
These explicit expressions satisfy the modular anomaly equation 
expressing $\mc S$-duality formulated in 
\cite{Billo:2013fi,Billo:2013jba,Billo:2014bja,Billo:2015ria,Billo:2015jta,Billo:2016zbf}.
Besides, by applying the AGT  correspondence, it is possible to obtain particular 
toroidal blocks  in closed form at  specific
values of the central charge $c$ and of the inserted operator conformal dimension.
Similarly, the  perturbative part of the prepotential can also be given in closed form 
and provides interesting special cases of the 3-point DOZZ Liouville 
correlation function \cite{Dorn:1994xn,Zamolodchikov:1995aa,Teschner:2003en}. 
In \cite{Beccaria:2016nnb}, the  complete list of all the 
$N\le 4$ poles points has been given, 
{\em i.e.} a set of  4, 7, 12, and 11 solutions at $N=1, 2, 3,4$ respectively.
Recently, these exact partition functions have been analyzed from the perspective of the 
Zamolodchikov's recursion relation for toroidal and spherical conformal blocks  in \cite{Nemkov:2016udj}.
This approach reveals once more the non-triviality of the solutions. In particular, 
the all order resummation of the instanton expansion in terms of 
quasi-modular forms remains an open question from the purely CFT point of view.
 
\medskip
Building on these results, in this paper we extend the analysis to the 
 $N_{f}$ theory with four hypermultiplet in the fundamental representation.\footnote{
A complementary analysis has been performed in \cite{Fucito:2013fba}
on a  class of $\mc N = 2$ gauge theories on $S^{4}$ with gauge group $U(N)$, and $N$ 
fundamental plus $N$ anti-fundamental hypermultiplets. 
It has been shown that for a specific values of the masses (being linear combinations of 
$\eps_{i}$ and $a$), the partition function is non trivial only at certain values of $a$ so that the 
full partition function $\int da |Z(a)|^{2}$ is given in terms of a finite number of residues
and can be explicitly evaluated by generalized hypergeometric functions. From the AGT perspective, these configurations are associated with four point correlators involving the insertion of a degenerated field. Our setup is quite different. Our hyper masses are proportional to linear combinations 
of the $\eps_{1,2}$ parameters only and the instanton partition functions will be a 
non-trivial function of $a$. Using AGT, the associated conformal blocks will have a generic intermediate conformal dimension and well defined modular properties.
} Indeed, as we discuss in Sec.~(\ref{sec:nek4}), the 
 explicit results obtained in \cite{Billo:2013fi} 
 and our previous experience in the $\mc N=2^{*}$ theory suggest the possibility of obtaining 
again closed partition functions in terms of modular Eisenstein series plus additional contributions 
involving two Jacobi theta functions. To illustrate our results, we begin the discussion by 
presenting in great details two special one-pole and one two-poles partition functions.
The first example has central charge $c=1$ (in the dual AGT conformal theory) and it is discussed in Sec.~(\ref{sec:ramond}). The mass and $\bm{\eps}$ parameters are constrained by the relations
\be
\label{1.3}
\frac{m_{1}}{\eps_{1}} = \frac{m_{3}}{\eps_{1}} = \frac{1}{2}, \qquad
\ \frac{m_{2}}{\eps_{1}} = \frac{m_{4}}{\eps_{1}} = 1, \qquad
\eps_{2} = -\eps_{1}.
\ee
An educated guess
based on explicit high instanton number computation reads
 \footnote{Here, $\E_{2}$ is the second Eisenstein series and
the dependence on $q$ is understood in all modular functions.}
\be
\label{1.4}
Z^{\text{inst}}(\bm{\eps}, a, \bm{m}; x)=\left( \frac{x}{16 \, q}  \right)^{-\frac{\nu^2}{4}} \left(1-x \right)^{-\frac{9}{4}} \vartheta_3^{-5}\, \left[1+\frac{1}{3 \nu^2}\left( - \E_2 + \vartheta_4^4 + 
2\, \vartheta_2^4\right)\right].
\ee
This expression may be read as  an all-order truncation of the results of  \cite{Billo:2013fi} at the special 
point (\ref{1.3}). To prove it, we can exploit AGT correspondence. As explained later, this amounts to the 
calculation of Virasoro conformal block in closed form with external conformal dimensions
$\Delta_{i} = ( \frac{1}{16}, \frac{9}{16}, \frac{9}{16}, \frac{1}{16})$, and internal 
{\em generic} dimension $\Delta = \frac{\nu^{2}}{4}$. This block can be computed 
in the Ramond sector of a massless scalar field and has been fully worked out 
by Zamolodchikov and Apikyan in \cite{Apikyan:1987mn}, generalizing the 
celebrated determination of the block of primary fields of dimension $\Delta_{i}=
(\frac{1}{16},\frac{1}{16},\frac{1}{16},\frac{1}{16})$ and generic internal dimension $\Delta$
in \cite{zamolodchikov1986two}. The result confirms perfectly the guessed expression (\ref{1.4}).
%

\medskip
We remark that the calculation in \cite{Apikyan:1987mn} is standard but somewhat laborious, in particular
if one wants to extend it to other cases. It would be quite nice if it were possible to obtain results like
(\ref{1.4}) by directly solving the Zamolodchikov recursion for conformal blocks at special points like 
(\ref{1.3}). This happens to be quite non-trivial, according to the analysis of \cite{Nemkov:2016udj} 
that is plagued by important technical problems. In particular, the special combination of 
quasi-modular forms in (\ref{1.4}) is a fact that remains a conjecture within this approach -- 
although it is true according to  the monodromy approach of \cite{Apikyan:1987mn}.
We solve these difficulties and 
provide a new {\em effective} recursion that allows to compute the block in closed form.
Again, this may appear as a mere confirmation of the result in \cite{Apikyan:1987mn}, but the advantage of the
 effective recursion is that it may be applied to other situations and prove more complicated
 expressions extending (\ref{1.4}) to two-poles partition functions, 
 as explained later. \footnote{Related remarks may be 
 found in the study of Zamolodchikov recursion for superconformal models, see 
 \cite{Belavin:2006zr,Hadasz:2006qb,Hadasz:2007ns,Hadasz:2007nt,Hadasz:2008zz,
 Hadasz:2008dt,Hadasz:2009db,Suchanek:2010zz,Suchanek:2010kq,Hadasz:2012im}
 and the dissertation at the following \href{http://www.fais.uj.edu.pl/documents/41628/d4817b20-2a2e-48d8-b9f6-40876b4a296f}{link}.  
 } We also emphasize that arbitrary conformal blocks at $c=1$ are computed in principle by 
 generic Painlev\'e VI tau functions  \cite{Gamayun:2012ma}, see also 
 \cite{Gamayun:2013auu,Iorgov:2014vla,Balogh:2014vya,Bershtein:2014yia,Bershtein:2016aef}.
 Special Painlev\'e solutions (of Riccati, Picard, Chazy and algebraic type) are in correspondence with specific
 blocks. For instance, the $(\frac{1}{16},\frac{1}{16},\frac{1}{16},\frac{1}{16})$ conformal block is 
 obtained from the Picard solution of \cite{kitaev1998solutions} and it would be interesting to discuss
 the block associated with  (\ref{1.4}) in this perspective. \\
 
 In the special deformed point (\ref{1.3}), we show that it is possible to give also the perturbative part of the 
 prepotential in closed form. Thus, we obtain  the exact quantum prepotential as
\be
\label{1.5}
F(\bm{\eps}, a, \bm{m}) =  F\pert(\bm{\eps}, a, \bm{m})+F\inst(\bm{\eps}, a, \bm{m}),
\ee
where the perturbative and instanton parts are respectively ($\Lambda$ is a perturbative ultraviolet scale)
\begin{align}
\label{1.6}
& F\pert(\bm{\eps}, a, \bm{m})= \,  2 \,\eps_1^2 \log \frac{a}{\Lambda} - 4\,a^2 \log 2, \\
& F(\bm{\eps}, a, \bm{m})= 2 \,\eps_1^2 \log \frac{a}{\Lambda} +
\eps_1^{2}\,\log\left(1+\frac{\eps_1^2}{12 \,a^2}\left( - \E_2 + \vartheta_4^4 + 2\, \vartheta_2^4\right)
\right). \notag
\end{align}

\medskip
The second example that we illustrate in details 
has central charge $c=-2$ and is discussed in Sec.~(\ref{sec:lcft}). 
The mass and $\bm{\eps}$ parameters are constrained by the relations
\be
\label{1.7}
\frac{m_{1}}{\eps_{1}} = \frac{1}{2}, \quad
\frac{m_{2}}{\eps_{1}} =
-\frac{m_{4}}{\eps_{1}} = \frac{1}{4}, \quad
\frac{m_{3}}{\eps_{1}} = 1, \quad
\eps_{2} = -\frac{1}{2}\,\eps_{1}.
\ee
Our guess is in this case 
\be
\label{1.8}
Z^{\text{inst}}(\bm{\eps}, a, \bm{m}; x)=\left( \frac{x}{16\, q}  \right)^{-\frac{\nu^2}{2}} \left(1-x \right)^{-2} \vartheta_3^{-5}\, \left[1+\frac{  -3 + 2 \E_2 + 2 \vartheta_2^4 
+ \vartheta_4^4 }{3\,(1 - 4 \nu^2)}\right].
\ee
and it matches several terms of the small instanton expansion. Now, 
the associated conformal block has external dimensions 
$\Delta_{i} = (-\frac{3}{32},\frac{5}{32},\frac{5}{32},\frac{21}{32})$ and {\em generic}
internal dimension $\Delta = \frac{1}{8}\left( 4\nu^2 -1 \right)$.
It may be realized in the $\mathbb Z_{4}$ orbifold sector of the $(\eta,\xi)$ ghost logarithmic CFT 
considered in \cite{Saleur:1991hk}. Many 4-point functions are available for such theories, 
see for instance \cite{Flohr:1995ea,Gaberdiel:1998ps,Kausch:2000fu,bhaseen2001disordered,
Flohr:2001zs,Kogan:2001nj,Pearce:2014sla}.
However, our case in (\ref{1.8}), and with generic internal dimension $\Delta$, is new.
Nevertheless, we are able to  compute it by applying the effective recursion mentioned above.
The result is in full agreement with the 
Ansatz (\ref{1.8}).
As in the previous $c=1$ case, the perturbative part can be computed exactly, and 
we are able to  give a closed formula
for the full prepotential
\be
\label{1.9}
F(\bm{\eps}, a, \bm{m}) = \,\eps_1^2 \log \frac{a}{\Lambda} +\frac{1}{2} \eps_1^2 \log \left(1-\frac{\eps_1^2 \left(2 \left(\E_2+\vartheta_2^4\right)+\vartheta_4^4\right)}{48\, a^2}\right).
\ee
In the same Sec.~(\ref{sec:lcft}) we also discuss a more complex example, where the scaling relations 
among the $\bm \eps$ and the $\bm m$ lead to a two-poles Nekrasov partition function. While it is slightly more complex
from the technical point of view, we can again provide an exact form for the partition function and the complete prepotential
\begin{align}
\label{1.10}
F(\bm{\eps}, a, \bm{m}) =&  \,2\, \eps_1^2 \log \frac{a}{\Lambda} + \frac{1}{2} \eps _1^2 \log \left( 1 + \frac{\eps _1^2 \left(-4\, \E_2 +6 \vartheta _2^4+3 \vartheta _4^4\right)}{16\, a^2}  + \right.\\ 
& \qquad \left.  \frac{\eps _1^4 \left(4\, \E_2^2-12\, \E_2 \vartheta _2^4+
\vartheta _4^4 \left(11 \vartheta _2^4-6\, \E_2\right)+11 \vartheta _2^8
+2 \vartheta _4^8\right)}{768\, a^4} \right).\notag
\end{align}  

\medskip
In Sec.~(\ref{sec:1P}), the previous examples are generalized and we provide a full list of one-pole 
Nekrasov partition functions with their associated exact conformal blocks. They are 
characterized by special values of the 5  ratios 
$\bm{\alpha}=\bm{m}/\eps_{1}$ and $\beta=\eps_{2}/\eps_{1}$.
The general form of the one-pole instanton partition function is 
\be
\label{1.11}
Z ^{\text{inst}}=\left( \frac{x}{16 q}  \right)^{\frac{\nu^{2}}{4\beta}} \left(1-x \right)^{ \frac{\left(\beta +1+\sum_{i=1}^4 \alpha_i\right){}^2}{4 \beta }}\,
 \vartheta_3(q)^{\frac{-(\beta +1)^2 + 2 \sum_{i=1}^4 \alpha_i^2}{\beta }}\, H_{\nu}(q),
\ee
where  
\be
\label{1.12}
H_{\nu}^{\text{1-pole}}(q) = \frac{\nu ^2+\frac{1}{\beta }\left( c_1(\bm{\alpha}, \beta) \,\E_2 + c_2(\bm{\alpha}, \beta) \,\vartheta_2^4 + c_3(\bm{\alpha}, \beta) \,\vartheta_4^4     \right)}
{\nu^{2}-\nu_1^{2}},
\ee
with certain explicit coefficients $c_{1,2,3}(\bm{\alpha}, \beta)$. The relevant 1-pole sets 
$\Pi_{1}=\{(\bm{\alpha}, \beta)\}$ contain 29 elements,
up to additional discrete transformations to be discussed later. In each case, 
the perturbative prepotential can be given in closed form and the full quantum prepotential 
turns out to be given by the following compact expression (this is (\ref{6.11}) copied here for the sake of 
presentation)
\begin{align}
\label{1.13}
F &= F\pert + F\inst =  -\frac{1}{2} \eps_1^2 \left(\beta(\beta+1)
 -2\,\sum_{i}\alpha_{i}^{2}+1\right) \log\left( \frac{a}{\Lambda} \right)  \\
& - \beta\,  \eps_1^2 \log \left(1+ \frac{\eps_1^2}{4\, a^2 \beta }\left( c_1(\bm{\alpha}, \beta) \,\E_2 + c_2(\bm{\alpha}, \beta) \,\vartheta_2^4 + c_3(\bm{\alpha}, \beta) \,\vartheta_4^4     \right)\right).\notag
\end{align}
The prepotential (\ref{1.13}) may be checked to pass a non trivial test, {\em i.e.}
the $\mc S$-duality  modular anomaly equation  
\cite{Billo:2013fi,Billo:2013jba,Billo:2014bja,Billo:2015ria,Billo:2015jta,Billo:2016zbf}.
All these solutions reduce to 
the massless $N_{f}=4$ theory in the undeformed limit $\bm{\eps}\to 0$. Nevertheless, in the deformed theory, they are non-trivial exact special partition functions.

\medskip
Finally, in Sec.~(\ref{sec:2P}), we present similar results for 2-pole solutions. In this case
we find a set $\Pi_{2}$ with 74 elements. The instanton partition function takes again the form (\ref{1.11}), but with a function $H_{\nu}^{\text{2-poles}}(q)$ fully discussed in (\ref{6.13})
and involving higher degree modular forms. Again, the perturbative prepotential can be computed
in closed form and nicely combine with the instanton contribution.

\medskip
To conclude, the expression (\ref{1.13}) and the analogous ones for the two-pole solutions, 
see  Sec.~(\ref{sec:2P}),
together with the full data characterizing the sets $\Pi_{1}$ and $\Pi_{2}$
are the main result of our paper. Once again, we remark that they can be proved 
by applying the novel tool of the effective recursion relation discussed in  
Sec.~(\ref{sec:effrec}).

\section{Exact partition functions in the deformed $\mc N=2^{*}$ theory}
\label{sec:review}

In this section we briefly summarize the results of  \cite{Beccaria:2016nnb} for the 
$\Omega$-deformed $\mathcal N=2^*$ $SU(2)$ gauge theory. Exact partition functions are found 
at particular values of the ratios
\be
\label{2.1}
\alpha = \frac{m}{\eps_{1}}, \qquad \beta = \frac{\eps_{2}}{\eps_{1}},
\ee
where $m$ is the mass of the hypermultiplet transforming in the adjoint representation.
We can parametrize the scalar field expectation value by setting $a = \frac{\nu}{2}\,\eps_{1}$.
The partition function is then 
\begin{equation}\label{2.2}
Z \left(\eps_1,\beta\, \eps_1,\frac{\nu}{2}\,\eps_1,\alpha\,\eps_1\right)=Z \left(1,\beta ,\frac{\nu}{2},\alpha\right)=\widetilde Z_{(\alpha,\beta)}(\nu).
\end{equation}
where we exploited the dimensional scaling indipendence to remove $\eps_1$
and we use a tilde to emphasize the new variables. We also define 
\begin{equation}\label{2.3}
\widetilde F_{(\alpha,\beta)}(\nu)= -\beta\log \widetilde Z_{(\alpha,\beta)}(\nu),
\end{equation}
which is the deformed prepotential that can be decomposed in 
its classical, perturbative, and instanton contributions
\begin{equation}\label{2.4}
\widetilde F_{(\alpha,\beta)}(\nu)=\widetilde F^{\text{class}}_{(\alpha,\beta)}(\nu)+\widetilde F^{\text{pert}}_{(\alpha,\beta)}(\nu)+\widetilde F^{\text{inst}}_{(\alpha,\beta)}(\nu).
\end{equation}
The main claim of \cite{Beccaria:2016nnb} is that 
there exists a finite set of $N$-poles points $(\alpha, \beta)$ such that 
the $k$-instanton prepotential is a rational function of $\nu$ with  poles at a fixed set of 
positions $\nu\in\{\nu_{1}, \dots, \nu_{N}\}$ independent on $k$.
This claim is definitely non-trivial and it has important consequences.
At the special $N$-poles points, it has been shown that the instanton
partition function and the perturbative part of the prepotential read, see App.~(\ref{sec:ids})
\be
\la{2.5}
\begin{split}
\wt Z\inst_{(\alpha, \beta)}(\nu) &= 
\frac{\nu^{2N}+\sum_{n=1}^{N}\nu^{2\,(N-n)}\mc M_{2n}(q)}{(\nu^{2}-\nu^{2}_{1})
\dots(\nu^{2}-\nu^{2}_{N})}\,[q^{-\frac{1}{12}}\,\eta(\tau)]^{2\,(\Delta_{m}-1)},\\
\wt F\pert_{(\alpha, \beta)}(\nu) &= -\beta\,\Delta_{m}\,\log\frac{\nu}{\Lambda}-\beta\log
\prod_{n=1}^{N}\bigg(1-\frac{\nu_{n}^{2}}{\nu^{2}}\bigg),
\quad \Delta_{m} = \frac{(\beta+1)^{2}-4\,\alpha^{2}}{4\,\beta}, 
\end{split}
\ee
where $\mc M_{2n}$ is a polynomial in the Eisenstein series $\E_{2}, \E_{4}, \E_{6}$
with  modular degree $2n$ and coefficients depending on $\alpha, \beta$, 
and $\Delta_{m}\in \mathbb N$. The quantity $\Lambda$ is the Seiberg-Witten UV scale.
The total prepotential is therefore 
remarkably simple and reads
\be
\la{2.6}
\wt F_{(\alpha, \beta)}(\nu) = -\beta\,\Delta_{m}\,\log\frac{\nu}{\Lambda}-\beta\,\log\bigg(
1+\sum_{n=1}^{N}
\frac{\mc M_{2n}(q)}{\nu^{2n}}\bigg).
\ee
The properties of this result have been discussed in the introduction. Here, we just add that 
according to AGT duality, the instanton partition function is related to the 1-point torus 
conformal block $\mathcal{F}_{\Delta_m}^{\Delta}$ by the relation \cite{Alday:2009aq} \footnote{
We remind that the torus conformal block 
$\mathcal{F}_{\Delta_m}^{\Delta}(q)$ is the trace of the {\em external} 
primary $\mc O_{\Delta_{m}}$ over the descendants of $\mc O_{\Delta}$ with a suitable $q$-dependent weight.
}
\begin{equation}\label{2.7}
Z^{\text{inst}}(q,a,m)= \left[\prod _{k=1}^\infty(1-q^{2k})\right]^{2\,\Delta_m-1}
\mathcal{F}_{\Delta_m}^{\Delta}(q).
\end{equation}
Here, the conformal dimension of the internal operator of the block is 
\begin{equation}\label{2.8}
\Delta=\frac{(\beta+1)^2-\nu^2}{4\beta},
\end{equation}
and gives the dependence on the vacuum expectation value $a$ appearing in $\nu$, see (\ref{2.2}).
Finally, the central charge is
\begin{equation}\label{2.9}
c=13+6\left(\beta+\frac{1}{\beta}\right).
\end{equation}
The four 1-pole solutions found in \cite{Beccaria:2016nnb} are associated with the 
CFT data $(c=0, \Delta_m=2)$ and $(c=-2, \Delta_m=3)$. 
The prediction for the conformal block is then 
\begin{equation}\label{2.10}
\mathcal{F}_{\Delta_m}^{\Delta}(q)=\frac{q^{\frac{1}{12}}}
{\eta(\tau)}\left[1+\frac{c-1}{24\,\Delta}(\text E_2 (q)-1)\right],
\end{equation}
with similar results  for the 2- and 3-pole solutions. In the rest of the paper, we shall generalize such 
kind of results to the $N_{f}=4$ theory. Here, AGT predicts exact expressions for spherical conformal 
blocks with four external operators and an exchanged internal operator with 
generic conformal dimension. \footnote{
Of course, in a definite CFT the internal operator is constrained by the fusion algebra, but the conformal block
may be defined {\em off-shell} and as such it is the object appearing in AGT correspondence.}

\section{Partition function and prepotential in the deformed $N_{f}=4$ theory}
\label{sec:nek4}

After the brief review of the $\mathcal{N}=2^{*}$ case, 
we now move to  the $\Omega$-deformed $\mathcal{N}=2$ SYM theory with gauge group $SU(2)$ and four flavour hypermultiplets in the fundamental representation. 
For non-vanishing masses of the hypermultiplets, the classical prepotential of the theory receives both  perturbative (1-loop) and  non-perturbative (instanton) quantum corrections
\be
\label{3.1}
F = F^{\text{pert}}+F^{\text{inst}},
\ee
where the instanton contribution  is related to the Nekrasov
partition function by the well known relation (\ref{1.2}) that we repeat here for the reader's convenience
\be
\la{3.2}
F\inst(\bm{\eps}, a, \bm{m})= -\eps_1\eps_2\,\log Z\inst(\bm{\eps}, a, \bm{m}).
\ee
The perturbative part instead can be expressed in terms of the generalized 
Barnes $\Gamma$ functions \cite{Nekrasov:2002qd,Nekrasov:2003rj}. The precise definition is the following.
Given the function
\be
\label{3.3}
\gamma_{\eps_1, \eps_2}(y) = \frac{d}{d s}\left( \frac{\Lambda^s}{\Gamma(s)} \int_{0}^{\infty} \frac{1}{t} \frac{t^s\,e^{- t y}}{\left( e^{- \eps_1 t} -1  \right)\left( e^{- \eps_2 t} -1  \right)} d t    \right) \Bigg|_{s=0},
\ee
the perturbative part of the prepotential  is 
\begin{align}
\label{3.4}
F\pert(\bm{\eps}, a, \bm{m})=&\,\eps_1 \eps_2 \left[  f(a) + f(-a) \right],
\end{align}
with 
\be
\label{3.5}
f(a)=\, \gamma_{\eps_1, \eps_2}(2\, a) - \sum_{f=1}^4{   \gamma_{\eps_1, \eps_2}\left(a + m_f + \frac{\eps_+}{2}\right) },
\ee
where we used the shorthand notation $\eps_{+} = \eps_1 + \eps_2$\footnote{We warn the reader that there are  different conventions in the literature
on the parametrization of the hypermultiplets masses, differing by a sign choice for $m_f$ and/or shift of the masses by $(\eps_1+\eps_2)/2$. 
Our conventions are the same, for example, of \cite{Billo:2013fi}.}. 
As anticipated in the introduction, our exact results for the Nekrasov partition functions (and consequently for the prepotential)
will involve simple combinations of modular and Jacobi theta functions.
Similar structures appear in the large $a$ expansion of the prepotential analysed in  \cite{Billo:2013fi}. We recall here explicitly some
of the results of  \cite{Billo:2013fi}, since they will be useful in the following sections for a direct comparison.
The complete quantum prepotential can be arranged in a series expansion in inverse powers of $a$ 
of the form
\be
\label{3.6}
F\inst(\bm{\eps}, a, \bm{m})+F\pert(\bm{\eps}, a, \bm{m})= 
h_0 \log \frac{a}{\Lambda} - \sum_{k=1}^{\infty}{\frac{h_k}{2^{k+1}\, k} \frac{1}{a^{2 k}}}.
\ee
The logarithmic term comes from the perturbative part, and both the perturbative and instanton parts contribute to the coefficients $h_k$. 
Rewriting the expansion (\ref{3.6}) 
in terms of $q$ instead of $x$, the authors of \cite{Billo:2013fi} managed to reconstruct the complete $q$ dependence 
of the first coefficients $h_k$, \emph{i.e.} their results encode the complete all-instanton contribution. 
The result is that the  functions $h_k$ can be expressed as  combinations of 
modular forms $\E_{2}, \E_4, \E_6$ and Jacobi theta functions $\vartheta_2$, $\vartheta_4$, 
with  $\eps_i$ and $m_f$ dependent polynomial coefficients. Explicitly, the first cases are
\begin{align}
 \label{3.7}	
h_0 = &\frac{1}{2} \left(4 R-\eps_{+} ^2+\eps_1 \eps_2\right),\notag\\
h_1 = &\frac{1}{24} \E_{2} \left(4 R-\eps_{+} ^2+\eps_1 \eps_2\right) \left(4 R-\eps_{+} ^2+3 \eps_1 \eps_2\right)-4 \left(T_1 \vartheta_4 {}^4-T_2 \vartheta_2 {}^4\right),\notag\\
h_2 = &\frac{1}{144} \E_{2} {}^2 \left(4 R-\eps_{+} ^2+\eps_1 \eps_2\right) \left(4 R-\eps_{+} ^2+3 \eps_1 \eps_2\right) \left(4 R-\eps_{+} ^2+4 \eps_1 \eps_2\right) +\notag\\
&\frac{4}{3} \E_{2} \left(4 R-\eps_{+} ^2+4 \eps_1 \eps_2\right) \left(T_2 \vartheta_2 {}^4-T_1 \vartheta_4 {}^4\right) + \notag\\
&\frac{1}{720} \E_4  \left[2304 N+2 \eps_1 \eps_2 \left(2 \left(80 R^2-88 R \eps_{+} ^2+17 \eps_{+} ^4\right)+\eps_1 \eps_2 \left(98 R-47 \eps_{+} ^2+15 \eps_1 \eps_2\right)\right)+
\right.\notag\\ 
& \left.\left(4 R-13 \eps_{+} ^2\right) \left(\eps_{+} ^2-4 R\right)^2\right] - \notag\\
&\frac{8}{3} \left(R-\eps_{+} ^2+\eps_1 \eps_2\right) \left(T_2 \vartheta_2 {}^8+2 \left(T_1+T_2\right) \vartheta_4 {}^4 \vartheta_2 {}^4+T_1 \vartheta_4 {}^8\right),
\end{align}
where $R, T_1, T_2$ and $N$ are given by the following $SO(8)$ invariant combinations of the masses
$\bm{m}$
\begin{align}
\label{3.8}
&R = \frac{1}{2} \sum_f m_f^2,\notag\\
&T_1=\frac{1}{12} \sum_{f<f'}{ m_f^2  m_{f'}^2}- \frac{1}{24} \sum_f m_f^4,\notag\\
& T_2 = -\frac{1}{24}\sum_{f<f'}{ m_f^2  m_{f'}^2} + \frac{1}{48}  \sum_f m_f^4 - \frac{1}{2} m_1 m_2 m_3 m_4,\\
& N = \frac{3}{16} \sum_{f<f'<f''}{ m_f^2  m_{f'}^2 m_{f''}^2} - \frac{1}{96}\sum_{f \neq f'} m_f^2 m_{f'}^4 + \frac{1}{96} \sum_f m_f^6.\notag
\end{align}
Starting from $h_3$, the expressions involve also the modular form $\E_6$. Note that the combination of masses
defined in (\ref{3.8}) transform non trivially under the action of the modular group \cite{Seiberg:1994aj}.
Taking into account  the modular properties of the  Jacobi theta functions, these transformations ensure that each $h_k$ transforms as a modular form
of degree $2k$, up to the anomalous term coming from  $\E_2$, see \cite{Billo:2013fi} for details.

Compared to the previous expansion, the meaning of the result (\ref{1.11}) is that it is possible to choose
special  ratios $m_{i}/\eps_{1}$ and $\eps_{2}/\eps_{1}$ in such a way that all the higher degree modular forms and 
powers of Jacobi theta functions in the prepotential are generated by the expansion of the 
logarithm of a finite number of 
terms in the partition function.


%

\subsection{Instanton partition function}
\label{sec:neknek}

The instanton partition function in the $\Omega$-deformed $\mathcal{N}=2$ SYM theory 
can be expanded order by 
order in the number of instantons according to 
\be
Z^{\text{inst}}(\bm{\eps}, a, \bm{m}) =1 +  \sum_{k=1}^{\infty} Z_k(\bm{\eps}, a, \bm{m})\, x^{k}.
\label{3.9}
\ee
As we discussed in the Introduction, it is convenient to trade $x$ by the modular parameter $q$
by means of the inverse of the relation (\ref{1.1}) 
\be
x=\frac{\vartheta_2^{4}(q)}{\vartheta_3^{4}(q)}.
\label{3.10}
\ee
We will refer to the $Z_k$ in expansion (\ref{3.9}) as
the \emph{Nekrasov functions}. At fixed instanton number $k$, the Nekrasov function can be calculated as a sum over pairs of $U(k)$ Young 
diagrams $(Y_1, Y_2)$, such that the total number of boxes $|Y_1|+ |Y_2| = k$ \cite{Alday:2009aq}.
To give a glimpse of the results and fix the notation, 
the simplest function $Z_1$ is given by the sum of the two terms
\begin{align}
 \label{3.11}
& Z_{(\Box,\,\bullet)} = \frac{\prod_{f=1}^4{\left(2 \left(a-m_f\right)-\eps_{+} \right)}}
{32\, a \,\eps_1 \,\eps_2 (\eps_{+} -2 a)},\notag \\
& Z_{(\bullet,\,\Box)} = -\frac{\prod_{f=1}^4{\left(2 \left(a+m_f\right)+\eps_{+} \right)}}
{32\, a \,\eps_1 \,\eps_2 (2 a+\eps_{+} )},\\
& Z_1(\bm{\eps}, a, \bm{m})=Z_{(\Box,\,\bullet)}+Z_{(\bullet,\,\Box)}.  \notag
\end{align}
Increasing $k$, the expressions for the Nekrasov functions become quickly cumbersome, but the structure of the $Z_k$ is always a rational function of the parameters, {\em i.e.},
\be
\label{3.12}
Z_{k}(\bm{\eps}, a, \bm{m}) = \frac{P_k(\bm{\eps}, a, \bm{m})}
{Q_k(\bm{\eps}, a, \bm{m})},
\ee  
where $P_k(\bm{\eps}, a, \bm{m})$ and $Q_k(\bm{\eps}, a, \bm{m})$ are polynomials in all the variables with 
increasing degree as $k$ grows.

\medskip

\subsection{Predictions  from AGT correspondence}

Before discussing AGT in this context, let us briefly summarize our notation for the conformal blocks
and recall Zamolodchikov's recursion.
Given a four point correlation function 
of primaries with dimensions $\Delta_i$ 
\be
\label{3.13}
\langle \mc O_{\Delta_4}(z_1) \mc O_{\Delta_3}(z_2) \mc O_{\Delta_2}(z_3) \mc O_{\Delta_1}(z_4) \rangle,
\ee
we can use a conformal transformation to set
$z_1=\infty, z_2 =1, z_3 = x$ and $z_4 = 0$, where $x$ now is the usual cross ratio 
$x=\frac{(z_1 - z_2)(z_3- z_4)}{(z_1-z_3)(z_2-z_4)}$.
The four point correlator can be expanded as a sum over intermediate states of dimension $\Delta$ 
\be
\label{3.14}
\langle \mc O_{\Delta_4}(\infty) \mc O_{\Delta_3}(1) \mc O_{\Delta_2}(x) \mc O_{\Delta_1}(0) \rangle = \sum_{\Delta}C_{\Delta_2,\Delta_1,\Delta}C_{\Delta,\Delta_3,\Delta_4}\,
\left|
x^{\Delta-\Delta_1-\Delta_2}\,
\confblock(B, \Delta, \Delta_{1}, \Delta_{2}, \Delta_{3}, \Delta_{4}, x) \right|^2\, ,
\ee
where   $C_{\Delta_i,\Delta_j,\Delta_k}$ are the 3-point structure constants. 
The function $B$ is the \emph{conformal block} and depends on the external dimensions $\Delta_i$, 
the internal dimension $\Delta$ and the central charge $c$ of the CFT (we omit this obvious dependence). \footnote{
It is convenient to strip off the explicit power of $x$ as in (\ref{3.14}) so that the small $x$ expansion of the 
conformal block start at one
\begin{equation*}
\confblock(B, \Delta, \Delta_{1}, \Delta_{2}, \Delta_{3}, \Delta_{4}, x)  = 1+
\frac{(\Delta-\Delta_1+\Delta_2)(\Delta+\Delta_3-\Delta_4)}{2 \Delta}\,x+\mc O(x^{2}).
\end{equation*}
}
It may be represented by the diagram
\begin{figure}[H]
\centering
\begin{tikzpicture}
 \node at (-1,1) {$\confblock(B, \Delta, \Delta_{1}, \Delta_{2}, \Delta_{3}, \Delta_{4}, x) \quad =$};
 \draw[line width=2pt, thick, black] (1,0.3) node [below, black] {${\Delta}_1$} -- ++(0.5,0)  -- ++(0.5,0) -- ++(0,1.5) 
 node[left, black] {${\Delta}_2$} -- ++(0,-1.5) -- ++(0.5,0) 
 node [below, black] {${\Delta}$} -- ++(0.5,0) -- ++(0,1.5) 
 node[right, black]{${\Delta}_3$} -- ++(0,-1.5) -- ++ (1,0) node [below, black] {${\Delta}_4$};
\draw +(6,0.75); 
\end{tikzpicture}
\end{figure}
\noindent
and admits the following \emph{elliptic} representation \cite{Zamolodchikov:1985ie}
\be
\label{3.15}
\confblock(B, \Delta, \Delta_{1}, \Delta_{2}, \Delta_{3}, \Delta_{4}, x) 
 = \left( \frac{x}{16 q}  \right)^{\frac{c-1}{24}-\Delta} \left(1-x \right)^{\frac{c-1}{24}-\Delta_2-\Delta_3} \vartheta_3(q)^{\frac{1}{2}\left(c-1- 8 \sum_i \Delta_i   \right)} 
 \confblock(H, \Delta, \Delta_{1}, \Delta_{2}, \Delta_{3}, \Delta_{4}, x) ,
\ee
where the relation between  $x$ and $q$ has been given in (\ref{3.10})
and the function $H$ obeys the Zamolodchikov's recursion relation \cite{Zamolodchikov:1985ie}. 
Using the following parametrization for the conformal dimensions in terms of the $\lambda_i$ variables
\be
\label{3.16}
\Delta_i = \frac{c-1}{24} + \lambda_i^2,  
\ee
and defining the quantities ($\Delta_{m,n}$ are the entries in the Kac table)
\begin{align}
\label{3.17}
& a_{\pm} = \frac{1}{\sqrt{24}}\left(  \sqrt{1-c} \pm \sqrt{25 -c}  \right),\quad  \lambda_{r,s} = a_{+} r + a_{-} s,  \\ 
& \Delta_{m,n}=\frac{c-1}{24}+\left(\frac{a_{+} m + a_{-} n}{2}\right)^2, \notag  \label{3.17}
\end{align}
the recursion formula for the function $H$ reads
\be
\label{3.18}
\confblock(H, \Delta, \Delta_{1}, \Delta_{2}, \Delta_{3}, \Delta_{4}, x) 
  = 1 + \sum_{m,n=1}^{\infty}\frac{ 16^{m n} q^{m n} R_{m,n}}{\Delta - \Delta_{m,n}} 
  \confblock(H, \Delta_{mn}+m\,n, \Delta_{1}, \Delta_{2}, \Delta_{3}, \Delta_{4}, x) ,
\ee
The coefficients $R_{m,n}$ appearing in (\ref{3.18}) are defined in terms of the $\lambda_i, \lambda_{r,s}$ introduced in (\ref{3.16}) and (\ref{3.17}) as follows 
\begin{align}
 \label{3.19}
 & R_{m,n} = \frac{P_{m,n}}{A_{m,n}},  \\
 & P_{m,n} = -\frac{1}{2} \prod_{r=-m+1,-m+3}^{m-1}\,\,\prod_{s=-n+1,-n+3}^{n-1} p_{r,s}, \qquad
A_{m,n}=\mathop{\prod_{k=-m+1}^{m} \prod_{l=-n+1}^n}_{(k,l)\neq(0,0),(m,n)}\lambda_{k,l},\notag\\
&p_{r,s} =\left( \lambda_1 + \lambda_2 - \frac{\lambda_{r,s}}{2} \right)\left( \lambda_2 - \lambda_1 -  \frac{\lambda_{r,s}}{2} \right) 
\left( \lambda_3 + \lambda_4 -  \frac{\lambda_{r,s}}{2} \right)\left( \lambda_3 - \lambda_4 -  \frac{\lambda_{r,s}}{2} \right). \notag
\end{align}
Equation (\ref{3.18}) is
understood as a sum over the residues of the conformal block in the poles $\Delta_{m,n}$. The products appearing in $ P_{m,n}$ are exactly 
those needed to cancel the contribution of null-states at the level $L = n m$ of the Verma module.


\subsubsection{The AGT dictionary for the $N_{f}=4$ theory}

The AGT correspondence states the equivalence between the Nekrasov partition function and a four point conformal block, 
identified through the following map between the gauge and CFT side parameters \cite{Alday:2009aq,Poghossian:2009mk}
\begin{align}
\label{3.20}
&\Delta = \frac{\eps_{+}^2}{4 \eps_1 \eps_2}- \frac{a^2}{\eps_1 \eps_2},\qquad
c=1+6 \frac{\eps_{+}^2}{\eps_1 \eps_2},\notag\\
&\Delta_i = \frac{\eta_i(\eps_{+} - \eta_i)}{\eps_1 \eps_2}, \qquad i=1\dots 4,\\
& \eta _1\ = \frac{1}{2} \left(\eps_{+}-m_1+m_2 \right),\qquad \eta_2 = \frac{1}{2} \left(\eps_{+} -m_1-m_2 \right),\notag\\
& \eta _3 = \frac{1}{2} \left(\eps_{+}-m_3-m_4 \right), \qquad \eta_4 = \frac{1}{2} \left(\eps_{+}-m_3+m_4 \right).	\notag
\end{align}
More precisely, the AGT statement is that the two functions
 $Z^{\text{inst}}(\bm{\eps}, a, \bm{m}; x) $ and  $B(\Delta, \Delta_{i}, c; x)$ are the same, 
using the above map, up to an overall $U(1)$ factor
\be
\label{3.21}
Z^{\text{inst}}(\bm{\eps}, a, \bm{m}; x)= (1-x)^{\kappa} \,
\confblock(B, \Delta, \Delta_{1}, \Delta_{2}, \Delta_{3}, \Delta_{4}, x)
\ee
with 
\be
\label{3.22}
\kappa =  \frac{2  (\eps_{+} - \eta_2)(\eps_{+} - \eta_3)}{\eps_1 \eps_2} = 
-\frac{\left(\sqrt{1-c+24 \Delta _2}+\sqrt{1-c}\right) \left(\sqrt{1-c+24 \Delta _3}+\sqrt{1-c}\right)}{12}.
\ee
Following \cite{Poghossian:2009mk}, it is convenient to rewrite the Nekrasov partition function in a form that mimics 
the elliptic form of the conformal block (\ref{3.15}). For compactness, we introduce the definition of the three exponents
\begin{align}
\label{3.23}
&X_1 = \frac{a^2}{\eps_1 \eps_2}, \qquad
X_2 = \frac{1}{4 \eps_1 \eps_2}\left( \eps_{+} + \sum_{f=1}^{4} m_f  \right)^2,\\
&X_3 = \frac{1}{\eps_1 \eps_2}\left( -\eps_{+}^2 + 2 \sum_{f=1}^{4}{m_f^2} \right).\notag
\end{align}
With these definitions, one can write \cite{Poghossian:2009mk}, see (\ref{3.15}),
\be
\label{3.24}
Z^{\text{inst}}(\bm{\eps}, a, \bm{m}; x)=\left( \frac{x}{16 q}  \right)^{X_1} \left(1-x \right)^{X_2} \vartheta_3(q)^{X_3}\, \confblock(H, \Delta, \Delta_{1}, \Delta_{2}, \Delta_{3}, \Delta_{4}, x) .
\ee
In the following sections  we are going to analyze  particular cases where the Nekrasov  functions have 
a very simple pole structure. In this context, the recursion (\ref{3.18})  has the clear advantage to 
make transparent the pole structure from the beginning. \\

The AGT dictionary is completed with the correspondence between the perturbative part of the  prepotential and the DOZZ formula for the 3-point function in the Liouville theory, see section 4.1 of \cite{Schiappa:2009cc}.

\section{The Ramond sector of scalar CFT and a special point with $c=1$}
\label{sec:ramond}

In this section we discuss a first example of a special point and we show  how the partition function 
can be obtained  in closed form. This is the point specified by the relations (\ref{1.3}) that we repeat 
here for convenience
\be
\label{4.1}
\frac{m_{1}}{\eps_{1}} = \frac{m_{3}}{\eps_{1}} = \frac{1}{2}, \qquad
\ \frac{m_{2}}{\eps_{1}} = \frac{m_{4}}{\eps_{1}} = 1, \qquad
\eps_{2} = -\eps_{1}.
\ee
Given the AGT map, the special point corresponds to the following set of external dimensions and central charge
\be
\label{4.2}
\Delta _1= \frac{1}{16},\quad
\Delta _2= \frac{9}{16},\quad
\Delta _3= \frac{9}{16},\quad
\Delta _4= \frac{1}{16},\quad c=1.
\ee
Alternative equivalent choices are possible because of the global $SO(8)$ flavour symmetry of the
gauge theory for $N_f=4$ and this symmetry is non trivial on the CFT side \cite{Giribet:2009hm}.
The choice in (\ref{4.2}) is particularly convenient for the following analysis.

\subsection{Perturbative contribution in closed form}

Given the relations (\ref{4.1}) the calculation of the perturbative part is straighforward. In fact for this special point, the perturbative part reduces to
\be
\label{4.3}
F\pert(\bm{\eps}, a, \bm{m}) = 2 \,\eps_1^2 \log \frac{a}{\Lambda} - a^2 \log 16.
\ee
In particular, this result means that all the coefficients $h_k$ with $k>0$ in (\ref{3.6}) are completely determined by the instanton contribution only.

\subsection{Instanton partition function at all orders}

A direct, brute force calculation of the Nekrasov partition function leads to the following result,
see Sec.~(\ref{sec:neknek})
\begin{align}
\label{4.4}
\widetilde{Z}^{\text{inst}}(\nu) &=  1+\sum_{k=1}^{\infty} \frac{1}{\nu^{2}}\,P_{k}(\nu)\,x^{k},
\end{align}
where the polynomials $P_{k}(\nu)$ have degree $2k+2$. The first polynomials are
\begin{align}
\label{4.5}
P_{1}(\nu) = &\frac{1}{8}\left(\,\nu ^4+13 \,\nu ^2+4\right),\notag\\
P_{2}(\nu) = &\frac{1}{256}\left(2 \,\nu ^6+65 \,\nu ^4+572 \,\nu ^2+288\right),\notag\\
P_{3}(\nu) = &\frac{1}{6144}\left(2 \,\nu ^8+117 \,\nu ^6+2383 \,\nu ^4+17484 \,\nu ^2+11232\right),\notag\\
P_{4}(\nu) = &\frac{1}{393216}\left(4 \,\nu ^{10}+364 \,\nu ^8+12771 \,\nu ^6+207023 \,\nu ^4+1361924 \,\nu ^2+1018176\right),\\
P_{5}(\nu) = &\frac{1}{15728640}\left(4 \,\nu ^{12}+520 \,\nu ^{10}+27555 \,\nu ^8+748070 \,\nu ^6+10527911 \,\nu ^4+
\right. \notag \\  &\qquad\qquad\qquad\qquad\qquad 
\left. 64320240 \,\nu ^2+53436240\right),\notag\\
P_{6}(\nu) = &\frac{1}{1509949440}\left(8 \,\nu ^{14}+1404 \,\nu ^{12}+104042 \,\nu ^{10}+4180545 \,\nu ^8+96592877 \,\nu ^6+ \right. \notag \\ 
&\qquad\qquad\qquad\qquad\qquad \left.
1232193096 \,\nu ^4+7132346928 \,\nu ^2+6409549440\right).\notag
\end{align}
Going to the function $H$ according to  (\ref{3.24}), 
we see that the  dependence on $\nu$ in the polynomials $P_{k}(\nu)$ is fully removed by the 
first factor in the r.h.s. of (\ref{3.24}), {\em i.e.} the power of the ratio 
$x/(16\,q)$ with a $\nu$-dependent exponent.
Including all the other factors appearing in the r.h.s. of (\ref{3.24}), we obtain 
\be
\label{4.6}
H_{\nu}(q)=1+\frac{8 \,q}{\nu ^2}+\frac{16\, q^2}{\nu ^2}+\frac{32\, q^3}{\nu ^2}+\frac{32\, q^4}{\nu ^2}+\frac{48\, q^5}{\nu ^2}+\frac{64\, q^6}{\nu ^2}+\mc O(q^7).
\ee
From this series  one may guess the following 
\emph{exact} $q$ dependence in terms of $\E_2$, $\vartheta_2$, and  $\vartheta_4$
\be
\label{4.7}
H_{\nu}(q)=1+\frac{1}{3 \nu^2}\left( - \E_2 + \vartheta_4^4 + 2 \vartheta_2^4\right).
\ee

\medskip

Upon the substitution $\Delta = \nu^2/4$, the exact form for $H$ in (\ref{4.7})
turns into a prediction for the
conformal block with conformal data as in (\ref{4.2}). In other words, our conjecture and AGT 
imply the remarkable result
\be
\label{4.8}
\confblock(H, \Delta, \frac{1}{16},\frac{9}{16},\frac{9}{16},\frac{1}{16}, x) =1+\frac{1}{12 \Delta}\left( - \E_2 + \vartheta_4^4 + 2 \vartheta_2^4\right).
\ee
In fact, as we mentioned in the Introduction, the relation (\ref{4.8}) has been obtained by Zamolodchikov
and Apikyan in \cite{Apikyan:1987mn} (see also \cite{Runkel:2001ng}). They considered
correlation functions in the Ramond sector of a free scalar field, which contains an infinite set of fields $\sigma_k$ with
dimensions $\Delta_k = (2 k +1)^2/16$. From the analysis of the operator product expansion 
they derived a set of operatorial relations for the $\sigma_k$
fields, and using these relations they were able to obtain, algorithmically, exact expressions for the conformal blocks. 
They also pointed out the simplicity of the pole structure of the conformal blocks. Among the explicit result presented in the paper, 
the formula (4.23) of  \cite{Apikyan:1987mn} is exactly the case under consideration here. 
Their result reads
\be
\label{4.9}
\confblock(H, \Delta, \frac{1}{16},\frac{9}{16},\frac{9}{16},\frac{1}{16}, x) = 1 - \frac{q}{\Delta} \frac{d}{d q} \log \vartheta_4(q) 
\ee
It is straightforward to use the properties of Jacobi functions in App.~(\ref{sec:ids}) to show 
that (\ref{4.8}) and (\ref{4.9}) are equivalent.

\subsection{The CFT side: a new method to derive exact conformal blocks}
\label{sec:effrec}

Exact expressions for the conformal blocks are rarely known. In this respect, the example 
(\ref{4.8}) considered in this section is a notable exception. 
In the rest of the paper we will extend the list of exact predictions from AGT, and in all the cases we will find similar 
structures in terms of the functions 
$\E_2$, $\vartheta_2$, and $\vartheta_4$.
 On the other hand, the method adopted in \cite{Apikyan:1987mn} is in principle applicable to other cases, but rather laborious.
It is then natural to investigate whether  this particular example
-- where everything is well under control both on the AGT and CFT side --  is characterised by  some hidden structure that allows for an exact calculation by means of  standard CFT techniques.

There are various approaches that allow the perturbative computation of the expansion of the conformal 
block in powers of the cross-ratio $x$ or of the elliptic parameter $q$. 
To give an example, the coefficient $B_k$ of  $x^k$ in the expansion of the conformal block can be derived using the so called 
hypergeometric recursion in the following form \cite{Perlmutter:2015iya} \footnote{For simplicity of notation, we do not write
explicitly the dependence of the coefficients $B_k(\Delta)$ on the external dimensions and the central charge.}
\begin{align}
\label{4.10}
& \confblock(B, \Delta, \Delta_1,\Delta_2,\Delta_3,\Delta_4, x) = \sum_{k=0} B_k(\Delta) x^k, \\
& B_k(\Delta) = \sum_{\ell=0}^k{\chi_\ell  \frac{(\ell + \Delta + \Delta_2 - \Delta_1)_{(k-\ell)}
(\ell + \Delta + \Delta_3 - \Delta_4)_{(k-\ell)}}{(k-\ell)! \,(2 \Delta +2 \ell)_{(k-\ell)}} },\notag
\end{align}
where $(a)_{n}$ is the Pochhammer symbol, and the coefficients 
$\chi_\ell$ are functions of the central charge, external and internal dimensions.
The  $\chi_\ell$ coefficients can be computed recursively and quite efficiently
as nested sums as explained in \cite{Perlmutter:2015iya} (see 
\cite{Beccaria:2015shq} for a recent application based on this
formalism). \footnote{
Another option is to evaluate the conformal block using the generalized Dotsenko-Fateev
(DF) 
representation according to the results of   \cite{Mironov:2010zs}. Technically, the hard
problem with this representation is the analytical continuation from integer DF exponents to rational values and this is ambiguous. Alternatively, one may exploit $SO(8)$ symmetry of the gauge theory to 
reduce to integer exponents, but then one needs to evaluate the DF integral with generic number of FD integration variables. The Jack polynomial machinery developed in 
\cite{Itoyama:2010ki,Maruyoshi:2014eja} is useful at low instanton numbers, but becomes increasingly unwieldy 
at higher orders and is ultimately equivalent to the Nekrasov determinant expressions and 
basically unuseful to obtain closed sums over all instantons.}
The first non trivial cases, at the CFT data in (\ref{4.1}), are
\begin{align}
\label{4.11}
\chi_0 &= 1, & \chi_1 &= 0, \notag  \\
\chi_2 &= \frac{1}{128} (2 \Delta +1), & 
\chi_3 &= \frac{3}{512 \Delta  \left(\Delta ^2+3 \Delta +2\right)},\\
\chi_4 &= \frac{(\Delta +2) \left(4 \Delta ^2+21 \Delta +23\right)}{32768 (\Delta +3)}, &
\chi_5 &=  \frac{3 (\Delta +8) (2 \Delta +9)}{65536 \Delta  (\Delta +2) (\Delta +3) (\Delta +4)}.\notag
\end{align}
Of course, using these $\chi_\ell$, one can reproduce the correct expansion of the complete block,
see (\ref{3.15})
\begin{align}
\label{4.12}
\confblock(B, \Delta, \frac{1}{16},\frac{9}{16},\frac{9}{16},\frac{1}{16}, x) = &1+\frac{(2 \Delta +1)^2 x}{8 \Delta }+\frac{(\Delta +2) (2 \Delta +1) (8 \Delta +9) x^2}{128 \Delta }+\notag\\
&\frac{(2 \Delta +3) \left(32 \Delta ^3+204 \Delta ^2+376 \Delta +153\right) x^3}{3072 \Delta }+\\
&\frac{(\Delta +4) \left(256 \Delta ^4+2496 \Delta ^3+8268 \Delta ^2+10403 \Delta+3822\right) x^4}{98304 \Delta }+\notag\\
&\frac{(2 \Delta +9) \left(512 \Delta ^5+8576 \Delta ^4+53688 \Delta ^3+152314 \Delta ^2+185395 \Delta +69990\right) x^5}{3932160 \Delta }+\notag\\
&\frac{1}{188743680 \Delta }(\Delta +6) (2 \Delta +9) \left(2048 \Delta ^5+40704 \Delta ^4+300704 \Delta ^3+ \right. \notag \\ 
&\left. \qquad\qquad\qquad\qquad\qquad 995460 \Delta ^2+1381229 \Delta +573180\right) x^6+\mathcal{O}\left(x^7\right)\notag .
\end{align}
Computationally the method is highly efficient, and one can push the above expansion to 
 high order.
On the other hand, it is hard to see any regularity or structure that may help in the search of the exact 
solution. From the previous discussion, these difficulties are somehow expected. Using the relations collected in appendix (\ref{sec:ids}), one can rewrite the exact expression 
of the block using only the $x$ variable as follows
\be
\confblock(B, \Delta, \frac{1}{16},\frac{9}{16},\frac{9}{16},\frac{1}{16}, x) =  \frac{\pi ^{5/2}\, 16^{\Delta }\,  x^{-\Delta } }{4 \sqrt{2} (1-x)^{9/8} \mathbb{K}(x)^{5/2}}  \left(1+ \frac{\mathbb{K}(x) (\mathbb{K}(x)-\mathbb{E}(x))}{\pi ^2 \Delta }\right) \left(e^{-\frac{\pi\,  \mathbb{K}(1-x)}{\mathbb{K}(x)}}\right)^{\Delta }.
\ee
This form makes clear that the dependence on $x$ of the block is rather intricate.

\subsubsection{Effective elliptic recursion}
\label{sec:effective}

A more promising strategy  is provided by the elliptic recursion. 
Following the results presented in \cite{Beccaria:2016nnb}, this approach to the problem was 
investigated in  \cite{Nemkov:2016udj}.
The solution of the recursion is in principle straightforward. Starting from  equation (\ref{3.18}) and 
expanding the function $H$ in series of $q$ as
\be
\label{4.14}
\confblock(H, \Delta, \Delta_{1}, \Delta_{2}, \Delta_{3}, \Delta_{4}, q)   = 1 + \sum_{k=1}^{\infty}{H_k(\Delta) q^k},
\ee
 one gets a recursive definition for the coefficients $H_k(\Delta)$
 \be
\label{4.15}
H_0(\Delta)=1,	\qquad
H_k(\Delta) = \mathop{\sum_{m,n\ge 1}}_{m n \le k}{\frac{16^{m\,n}\,R_{m,n}}{\Delta - \Delta_{m,n}} H_{k - m n}(\Delta_{m,n} + m n)  }.
\ee
The usual strategy, given the external dimensions, is to compute the $R_{m,n}$  leaving $c$ generic, to avoid 
poles in (\ref{3.19}), 
solve the recursion and finally take the limit $c\to1$. 
Following this route one recovers the correct series expansion, but the nested structure of the recursive solution (\ref{4.18}) is still not very transparent. On the other hand, one knows that at the point under consideration
the block has the quite simple dependence on $\Delta$ as in 
\be
\label{4.16}
\confblock(H, \Delta, \Delta_{1}, \Delta_{2}, \Delta_{3}, \Delta_{4}, q)  = 1+\frac{1}{\Delta} \mathcal{H}(\Delta_i; q).
\ee
If one starts at $c=1$ from the beginning in order to 
possibly truncate the recursion using (\ref{4.16}) in (\ref{4.15}) the problem is that of the 
reducibility of  Verma module. In other words, the equation 
\be
\label{4.17}
\Delta_{m,n} + m\, n = \Delta_{m', n'},
\ee
has integer solutions. This means that a naive application of (\ref{4.15}) leads to undetermined
ratios $0/0$ preventing truncation, see \cite{Nemkov:2016udj}.

\medskip
Here, we propose another strategy that overcomes these problems and is able to 
provide  the function $\mc H$ in (\ref{4.16}) in closed form with little effort. 
To illustrate the  basic idea  let us consider the case of the $R_{2,2}$ term. As a function of $c$, in the special point
one gets
\be
\label{4.18}
R_{2,2}(c)= \frac{1}{16^4}\frac{(c-27)^2 (c-1) ((c-12) c-37)^2}{256 (c-28) (c-25) (c+2)},
\ee
and clearly  $\lim_{c\to1} R_{2,2}(c)=0$. Instead, working with $c=1$, the arguments of the pro\-ducts $P_{m,n}$ and $A_{m,n}$ become $\frac{1}{16} (r-s+1)^2 (r-s+2)^2$ and $k - l$ respectively. 
Taking into account the range of the indices, for $R_{2,2}$ we have a "$0^2$" in the numerator from the pair $(r,s)=(-1,1)$, but also 
two factors vanishing in the denominator for $(k,l)=(-1,-1)$ and $(k,l)=(1,1)$. The final  $0^2/0^2$ expression is finite. This order of limits ambiguity has been clearly pointed out 
in  \cite{Nemkov:2016udj}\footnote{A similar problem is  discussed in appendix B of the 
dissertation at the following
 \href{http://www.fais.uj.edu.pl/documents/41628/d4817b20-2a2e-48d8-b9f6-40876b4a296f}{link}. 
 A theoretical framework for this kind of problems has been recently
formulated in \cite{Itoyama:2014dya}.}.
To evaluate practically the products  with $c=1$, we need some regularization. 
We propose  the simplest possible prescription, introducing a shift $\delta$
in one of the indices of the products and then taking the $\delta \to 0$ limit at the end. Our \emph{effective} definition of the $R_{m,n}$ coefficients is then based on the definitions
\begin{align}
\label{4.19}
& P_{m,n} = -\frac{1}{2} \prod_{r=-m+1,-m+3}^{m-1}\,\,\prod_{s=-n+1,-n+3}^{n-1} 
p_{r,s+\delta}, \qquad\qquad
A_{m,n}=\mathop{\prod_{k=-m+1}^{m} \prod_{l=-n+1}^n}_{(k,l)\neq(0,0),(m,n)}
\lambda_{k,l+\delta},
\end{align}
with (in our special  $c=1$ point)
\be
\label{4.20}
 p_{r,s+\delta} = \frac{1}{16} (-\delta +r-s+1)^2 (-\delta +r-s+2)^2,\qquad  \lambda_{k,l+\delta} =k-l-\delta.
\ee
At the end, we remove the regularization and compute the {\em effective} coefficients
 \be
 \label{4.21}
 R_{m,n}^{\text{eff}} = \lim_{\delta\to 0}\frac{P_{m,n}}{A_{m,n}}. 
\ee
The first advantage of this approach is that the coefficients $R_{m,n}^{\text{eff}}$ 
can be easily computed in closed form. They read
\be
\label{4.22}
R_{m,n}^{\text{eff}}=-\frac{2}{16^{m^2}} (-1)^m m^2  \delta_{m,n},
\ee
and, as expected, they are non vanishing only for $\Delta_{m,n} = 0$.
Plugging the Ansatz  (\ref{4.16}) into (\ref{4.15}) and using 
$R\to R^{\text{eff}}$, we get the {\em effective recursion relation} 
%
\be
\label{4.23}
1+\frac{1}{\Delta} \mathcal{H} = 1 - 2 \sum_{m=1}^{\infty}{\frac{(-1)^m \,m^{2}\, 
q^{m^2}}{\Delta} \left( 1+\frac{1}{m^2}  \mathcal{H}  \right)}.
\ee
We can solve (\ref{4.23}) for  $\mathcal{H}$ as the ratio of two series
\be
\label{4.24}
\mathcal{H}  = -\frac{2 \sum_{m=1}{(-1)^m m^2 q^{m^2}}}{1+ 2 \sum_{m=1}{(-1)^m  q^{m^2}}}.
\ee
The denominator is nothing but $\vartheta_4$, while  the numerator is its $q\,\partial_{q}$ .
In this way one recovers our result in the form of eq. (\ref{4.9}). Later, we shall show that this approach
is general and powerful providing similar simple computation of the special exact conformal blocks
appearing in our analysis.

\subsection{Closed form of the full prepotential}

Putting together the results for the perturbative and instanton parts  we can now write the full prepotential
\begin{align}
\label{4.25}
& F\pert(\bm{\eps}, a, \bm{m})= \,  2 \,\eps_1^2 \log \frac{a}{\Lambda} - a^2 \log 16, \\
& F\inst(\bm{\eps}, a, \bm{m})= \eps_1^{2}\,\log\left( \left( \frac{x}{16 \, q}  \right)^{-\frac{a^2}{\eps_1^2}} \left(1-x \right)^{-\frac{9}{4}} \vartheta_3^{-5}\, \left(1+\frac{\eps_1^2}{12 \,a^2}\left( - \E_2 + \vartheta_4^4 + 2\, \vartheta_2^4\right)\right)\right),\notag\\
& F(\bm{\eps}, a, \bm{m}) =  F\pert(\bm{\eps}, a, \bm{m})+F\inst(\bm{\eps}, a, \bm{m}).\notag
\end{align}
Expanding for $a \to \infty$, and discarding all the $a$ independent terms, we can compare directly with the 
the large $a$  expansion (\ref{3.6}).
The term $(\frac{x}{16 \, q})^{-a^2/\eps_1^2}$ in the instanton part
combines with the corresponding term proportional to $a^2$ in the perturbative part and 
gives the renormalization of the gauge
coupling constant \cite{Billo:2013fi}. The check for the first coefficient $h_1$ is trivial. 
Moreover, equation (\ref{4.25}) implies that \emph{all} the coefficients $h_k$ with $k>1$ must simply be powers of $h_1$, up to the numerical coefficient 
arising from the expansion of the logarithm. This is indeed what happens. 
The second coefficient for $a^{-4}$ arising from the expansion of (\ref{4.25}) is 
\be
\label{4.26}
-\frac{1}{288} \eps _1^6 \left(-\E_2+\vartheta^4+2 \vartheta _2^4\right){}^2,
\ee
which is in agreement with  (\ref{3.6}), evaluating $h_2$ at  the special point and using the identity 
$\E_4 = \vartheta _2^8+\vartheta _4^4 \vartheta _2^4+\vartheta _4^8$. 
The same check can be performed with $h_3$, and confirms again the pattern.

\section{Logarithmic CFT and examples with $c=-2$}
\label{sec:lcft}

In this section we present other two examples of special points where the Nekrasov partition function has a finite 
number of poles and can be worked out exactly. They are characterized by the scaling relations, see (\ref{1.7})
\be
\label{5.1}
\frac{m_{1}}{\eps_{1}} = \frac{1}{2}, \quad
\frac{m_{2}}{\eps_{1}} =
-\frac{m_{4}}{\eps_{1}} = \frac{1}{4}, \quad
\frac{m_{3}}{\eps_{1}} = 1, \quad
\eps_{2} = -\frac{1}{2}\,\eps_{1}.
\ee
and
\be
\label{5.2}
\frac{m_1}{\eps_1}=\frac{1}{2},\ \quad
\frac{m_2}{\eps_1}= 
\frac{m_3}{\eps_1}=\frac{3}{4},\ \quad \ \
\frac{m_4}{\eps_1}=1,\ \quad 
\eps_{2} = -\frac{1}{2}\,\eps_{1},
\ee
respectively. For the first one, the Nekrasov functions have again a single pole, for the second choice two poles. Through AGT they are related
to a logarithmic CFT with $c=-2$, as discussed in more detail below. \\

Before moving to the calculation of the instanton partition functions, we give here the result for the perturbative 
part of the prepotential. In both cases the calculation can be easily done exactly, and the result is
\be
\label{5.3}
F\pert(\bm{\eps}, a, \bm{m}) = \kappa \,\eps_1^2 \log \frac{a}{\Lambda} - a^2 \log 16 + \frac{1}{2} \,\eps_1^2 \log \left(1-\frac{\eps _1^2}{16\, a^2}\right),
\ee
with $\kappa=1$ for the first special point, and $\kappa=2$ for the second one. We remark that the perturbative 
part of the prepotential may be given in clean form in the undeformed massive limit, 
see for instance (2.16) of \cite{Billo:2015ria}. The simplicity of (\ref{5.3}) holding for non zero deformation
is remarkable and similar to the recent findings in \cite{Beccaria:2016nnb}.

\subsection{A one-pole point}

Given the scaling relations in (\ref{5.1}), the AGT map leads to the corresponding set of external dimensions 
\be
\label{5.4}
\Delta _1= -\frac{3}{32},\quad \Delta _2= \frac{5}{32},\quad
\Delta _3= \frac{5}{32},\quad \Delta _4= \frac{21}{32},
\ee
and the central charge of the CFT is $c=-2$. It is easy to recognize that the values in (\ref{5.4})
 appear
in the extended Kac table of the logarithmic minimal model for critical dense polymers, usually denoted by $\mathcal{L}(1,2)$. 
This is the first (and best studied)  example of a family  of integrable logarithmic minimal model $\mathcal{L}(p,p')$, with $1<p<p'$ and $p, p'$ coprimes, 
describing  non-intersecting loops in a lattice.
The CFT description corresponds to the $\mathbb Z_{4}$  sector of the $(\eta,\xi)$ ghost logarithmic CFT \cite{Saleur:1991hk}, usually referred 
as \emph{symplectic fermions} \cite{Kausch:2000fu}. The dimension $-\frac{3}{32}$ corresponds to the ground state of the $\mathbb Z_{4}$ sector.  
The states with dimensions $\frac{21}{32}$ and  $\frac{5}{32}$ are obtained
by acting on the ground state with a  single creation mode
of the $\eta$ and $\xi$ fields, respectively.\\
To discuss our results, we follow the same steps of the previous example. The calculation of the Nekrasov partition function gives
\begin{align}
\label{5.5}
\widetilde{Z}^{\text{inst}}(\nu) &=  1+\sum_{k=1}^{\infty} \frac{1}{4 \nu ^2-1}\,P_{k}(\nu)\,x^{k},
\end{align}
where now the first polynomials $P_{k}(\nu)$ are given by
\begin{align}
\label{5.6}
P_{1}(\nu) = & \frac{1}{8} \left(8 \,\nu ^4+42 \,\nu ^2-15\right), \notag\\
P_{2}(\nu) = &\frac{1}{128} \left(16 \,\nu ^6+224 \,\nu ^4+783 \,\nu ^2-330\right), \notag\\
P_{3}(\nu) = & \frac{32 \,\nu ^8+832 \,\nu ^6+7330 \,\nu ^4+20963 \,\nu ^2-9852}{3072}, \\
P_{4}(\nu) = & \frac{128 \,\nu ^{10}+5280 \,\nu ^8+82792 \,\nu ^6+580950 \,\nu ^4+1461705 \,\nu ^2-745290}{196608}, \notag\\
P_{5}(\nu) = & \frac{1}{7864320} \left( 256 \,\nu ^{12}+15296 \,\nu ^{10}+369200 \,\nu ^8+4494100 \,\nu ^6+27350124 \,\nu ^4+
\right. \notag \\ & \left.\qquad\qquad
62752839 \,\nu ^2-34153470 \right), \notag \\
P_{6}(\nu) = & \frac{1}{377487360}  \left( 512 \,\nu ^{14}+41728 \,\nu ^{12}+1426880 \,\nu ^{10}+26202400 \,\nu ^8+271958978 \,\nu ^6+
\right. \notag\\ & \left.\qquad\qquad
1497332707 \,\nu ^4+3199335720 \,\nu ^2-1838760300  \right).\notag
\end{align}
Again, once the prefactor is removed, the result simplifies drastically and the function $H$  has a very simple expansion
\be
\label{5.7}
H_{\nu}(q)=1-\frac{8\, q}{4 \nu ^2-1}+\frac{8\, q^2}{4 \nu ^2-1}-\frac{32\, q^3}{4\, \nu ^2-1}+\frac{40\, q^4}{4 \nu ^2-1}-\frac{48\, q^5}{4 \nu ^2-1}+\frac{32\, q^6}{4 \nu ^2-1}+\mathcal{O}(q^7),
\ee
suggesting the following exact form
\be
\label{5.8}
H_{\nu}(q)=1+\frac{3 - 2\, \E_2 - 2\, \vartheta_2^4 - \vartheta_4^4 }{3\,(4 \nu^2-1)}.
\ee
In this second example the relation between $\nu$ and the internal conformal 
dimension $\Delta$ is $\nu = \frac{1}{2} \sqrt{8 \Delta +1}$. Thus, equation (\ref{5.8})
is equivalent to the following prediction for the conformal block
\be
\label{5.9}
\confblock(H, \Delta, -\frac{3}{32},\frac{5}{32},\frac{5}{32},\frac{21}{32}, x) =1+\frac{1}{24 \Delta}\left(3 - 2 \E_2 - 2 \vartheta_2^4 - \vartheta_4^4  \right).
\ee
To prove (\ref{5.9}) by a direct CFT calculation we use the effective recursion presented in 
Sec.~(\ref{sec:effective}). 
Given the external dimensions and $c=-2$, the arguments of the products $P_{m,n}$ and $A_{m,n}$ are found to be
\be
\label{5.10}
 p_{r,s} = \frac{1}{64} (2 r-s-1) (2 r-s+1) (2 r-s+2) (2 r-s+4),\qquad  \lambda_{k,l} =\sqrt{2}\, k-\frac{l}{\sqrt{2}}.
\ee
The coefficients $R_{m,n}$  are calculated with the same regularization prescription
discussed in Sec.~(\ref{sec:effective}), \emph{i.e.} we shift one of the indices by $\delta$
to compute the products, and take  $\delta \to 0$ at the end.  The resulting 
{\em effective} coefficients $R_{m,n}^{\text{eff}}$ is again very simple (compare with (\ref{4.22}))
\be
\label{5.11}
R_{m,n}^{\text{eff}}= \begin{cases}
-\frac{1}{16^{m(2 m +1)}} m (2 m + 1), & \text{for  } n= 2 m +1 \\
-\frac{1}{16^{m(2 m -1)}} m (2 m - 1), & \text{for  }   n= 2 m -1 \\
0 & \text{otherwise.} 
\end{cases}
\ee
Inserting in the effective recursion the expression (\ref{5.11}) together with the Ansatz (\ref{4.16}),
we get  the following  equation for $\mathcal{H}$
\begin{align}
\label{5.12}
\mathcal{H} = & - \sum_{m=1}^{\infty}{m \left( 2 m + 1 \right) q^{m(2 m +1)} \left( 1+\frac{1}{m (2 m+1)}  \mathcal{H}  \right)} - \\
&   \sum_{m=1}^{\infty}{m \left( 2 m - 1 \right) q^{m(2 m -1)} \left( 1+\frac{1}{m (2 m-1)}  \mathcal{H} \right)},\notag
\end{align}
which gives
\be
\label{5.13}
\mathcal{H} =- \frac{\sum_{m=1}^{\infty}
\left[m \left( 2 m + 1 \right) q^{m(2 m +1)}+m \left( 2 m - 1 \right) q^{m(2 m -1)}\right]}
{1+ \sum_{m=1}^{\infty}\left[q^{m(2 m +1)} + q^{m(2 m -1)}\right] }.
\ee
Finally, summing the series, one recovers the combination $\frac{1}{24}\left(3 - 2 \E_2 - 2 \vartheta_2^4 - \vartheta_4^4  \right)$ in equation (\ref{5.9}), {\em i.e.},
\be
\label{5.14}
\mc H = \frac{1}{24}(3-2\,\E_{2}-2\vartheta_{2}^{4}-\vartheta_{4}^{4}).
\ee
To prove this fact, we begin by observing that the denominator of (\ref{5.13}) may be written as
\be
\label{5.15}
1+ \sum_{m=1}^{\infty}\left[q^{m(2 m +1)} + q^{m(2 m -1)}\right]  = \sum_{m=1}^{\infty}
q^{m(m-1)/2} = \frac{1}{2}\,q^{-1/8}\,\vartheta_{2}(\sqrt{q}).
\ee
Hence, we have 
\be
\label{5.16}
\mc H = -q \frac{d}{dq}\log\bigg(\frac{1}{2}\,q^{-1/8}\,\vartheta_{2}(\sqrt{q})\bigg).
\ee
Using (\ref{A.11}), this is 
\be
\label{5.17}
\mc H = -\frac{1}{2}\,q \frac{d}{dq}\log\bigg(\frac{1}{2}\,q^{-1/4}\vartheta_{2}(q)
\,\vartheta_{3}(q)\bigg).
\ee
The derivative can be computed using (\ref{A.10}). After simplification by (\ref{A.9})
we prove (\ref{5.14}).

\subsection*{Complete prepotential}

Our results for the perturbative and instanton contribution to the prepotential are
\begin{align}
\label{5.18}
F\pert(\bm{\eps}, a, \bm{m}) &= \,\eps_1^2 \log \frac{a}{\Lambda} - 4\,a^2 \log 2 + \frac{1}{2}\, \eps_1^2 \log \left(1-\frac{\eps_1^2}{16\, a^2}\right), \\
F\inst(\bm{\eps}, a, \bm{m}; x) &=\frac{1}{2} \eps_1^2  \log \left(  \left( \frac{x}{16\, q}  \right)^{-\frac{2 a^2}{\eps_1^{2}}} \left(1-x \right)^{-2} \vartheta_3^{-5}\, \left(1+\frac{  -3 + 2 \E_2 + 2 \vartheta_2^4 
+ \vartheta_4^4 }{3\,(1 - \frac{ 16 \, a^2}{\eps_1^2})}\right) \right).\notag
\end{align}
Summing them, we have again a closed form for $F = F\pert(\bm{\eps}, a, \bm{m})+F\inst(\bm{\eps}, a, \bm{m}; x)$.
In this case the perturbative part contributes to all the $h_k$ coefficient, and combining the logarithms one gets the very compact expression 
\be
\label{5.19}
F(\bm{\eps}, a, \bm{m}) = \,\eps_1^2 \log \frac{a}{\Lambda} +\frac{1}{2} \eps_1^2 \log \left(1-\frac{\eps_1^2 \left(2 \left(\E_2+\vartheta_2^4\right)+\vartheta_4^4\right)}{48\, a^2}\right).
\ee
As a further check, we have successfully tested (\ref{5.19}) by comparing it 
with the general large $a$ expansions, as we did in the previous example.

\subsection{A two poles point}
\label{sec:two-example}

Here we move to the second more complex example, where the Nekrasov partition function and its dual block have two poles. The scaling relations are the ones written in (\ref{5.2}), and they correspond
to the following external dimensions for the conformal block
\be
\label{5.20}
\Delta _1= -\frac{3}{32},\quad
\Delta _2= \frac{21}{32},\quad
\Delta _3= \frac{45}{32},\quad
\Delta _4= -\frac{3}{32},
\ee
again with central charge $c=-2$.
The complete partition functions is
\begin{align}
\label{5.21}
\widetilde{Z}^{\text{inst}}(\nu) &=  1+\sum_{k=1}^{\infty} \frac{1}{\nu^2(4 \nu ^2-1)}\,P_{k}(\nu)\,x^{k}.
\end{align}
The list of the first polynomials $P_{k}(\nu)$ in this case is given by
\begin{align}
\label{5.22}
P_{1}(\nu) = & \frac{1}{4} \,\nu ^2 \left(4 \,\nu ^4+79 \,\nu ^2-2\right), \notag\\
P_{2}(\nu) = &\frac{1}{128} \left(16 \,\nu ^8+688 \,\nu ^6+7687 \,\nu ^4+1323 \,\nu ^2+60\right), \notag\\
P_{3}(\nu) = & \frac{16 \,\nu ^{10}+1112 \,\nu ^8+26573 \,\nu ^6+221431 \,\nu ^4+75642 \,\nu ^2+4536}{1536} , \\
P_{4}(\nu) = & \frac{128 \,\nu ^{12}+12704 \,\nu ^{10}+485160 \,\nu ^8+8492902 \,\nu ^6+58818493 \,\nu ^4+28571244 \,\nu ^2+2117664}{196608}, \notag\\
P_{5}(\nu) = & \frac{1}{3932160}\left(128 \,\nu ^{14}+16928 \,\nu ^{12}+915240 \,\nu ^{10}+25357150 \,\nu ^8+362520277 \,\nu ^6
\right. \notag \\ & \left.\qquad\qquad
+2207662182 \,\nu ^4+1342390200 \,\nu ^2+117034560 \right),\notag\\
P_{6}(\nu) = & \frac{1}{377487360}\left( 512 \,\nu ^{16}+86272 \,\nu ^{14}+6171584 \,\nu ^{12}+240341920 \,\nu ^{10}+5392008818 \,\nu ^8+
\right. \notag \\ & \left.\qquad\qquad
66661116493 \,\nu ^6+369193795431 \,\nu ^4+262864732140 \,\nu ^2+26081676000\right).\notag
\end{align}
Removing as usual the prefactor, the resulting $H$ has the expansion
\begin{align}
\label{5.23}
H_{\nu}(q)=&1+\frac{72 q}{4 \nu ^2-1}+\frac{24 \left(7 \nu ^2+5\right) q^2}{\nu ^2 \left(4 \nu ^2-1\right)}+\frac{288 \left(\nu ^2+2\right) q^3}{\nu ^2 \left(4 \nu ^2-1\right)}+\frac{72 \left(5 \nu ^2+22\right) q^4}{\nu ^2 \left(4 \nu ^2-1\right)}+\\
& \frac{432 \left(\nu^2+8\right) q^5}{\nu ^2 \left(4 \nu ^2-1\right)}+\frac{672 \left(\nu ^2+9\right) q^6}{\nu ^2 \left(4 \nu ^2-1\right)}+\frac{576 \left(\nu ^2+18\right) q^7}{\nu ^2 \left(4 \nu ^2-1\right)}+\frac{24 \left(31 \nu ^2+644\right) q^8}{\nu ^2 \left(4 \nu^2-1\right)}+\notag\\
& \frac{72 \left(13 \nu ^2+320\right) q^9}{\nu ^2 \left(4 \nu ^2-1\right)}+\frac{1008 \left(\nu ^2+31\right) q^{10}}{\nu ^2 \left(4 \nu ^2-1\right)}+\frac{864 \left(\nu ^2+50\right) q^{11}}{\nu ^2 \left(4 \nu ^2-1\right)}+\mathcal{O}\left(q^{12}\right).\notag
\end{align}
Despite the structure of the expansion is clearly more involved than in the previous cases, one can guess again a closed form
\be
\label{5.24}
H_{\nu}(q)=\frac{\nu ^4-2 h_1 \nu ^2+2 \left(h_1^2-h_2\right)}{\nu ^2 \left(\nu ^2-\frac{1}{4}\right)},
\ee
where $h_1$ and $h_2$ are the coefficients in equation (\ref{3.7}) evaluated at the special point.
Explicitly they are:
\begin{align}
\label{5.25}
h_1 = &\frac{1}{8} \left(4 \E_2-6 \vartheta _2^4-3 \vartheta _4^4\right),\\
h_2 = &\frac{1}{192} \left(40\E_2^2+6 \vartheta _2^4 \left(7 \vartheta _4^4-20 \E_2\right)-60 \E_2 \vartheta _4^4+44 \E_4+42 \vartheta _2^8-21 \vartheta _4^8\right) . \notag
\end{align}
Moving to the CFT side using  $\nu = \frac{1}{2} \sqrt{8 \Delta +1}$, it is useful to rewrite the prediction for the conformal block as
\be
\label{5.26}
\confblock(H, \Delta, -\frac{3}{32},\frac{21}{32},\frac{45}{32},-\frac{3}{32}, x) =1+\frac{{H}_1}{8 \Delta +1}+\frac{{H}_2}{8 \Delta }.
\ee
where the two functions ${H}_1$ and ${H}_2$ are given by the following combinations of $\E_2$ and Jacobi theta functions:
\begin{align}
\label{5.27}
H_1 = &\frac{1}{3} \left(-4 \E_2^2+12 \E_2 \vartheta _2^4+\vartheta _4^4 \left(6 \E_2-11 \vartheta _2^4\right)-11 \vartheta _2^8-2 \vartheta _4^8\right),\\
H_2 = & \frac{1}{3} \left(\vartheta _2^4 \left(-12 \E_2+11 \vartheta _4^4+18\right)+(9-6 \E_2) \vartheta _4^4+4 (\E_2-3) \E_2+11 \vartheta _2^8+2 \vartheta _4^8+3\right) . \notag
\end{align}
The strategy to prove our guess is again to use the effective recursion.
In this case we have
\be
\label{5.28}
 p_{r,s} = \frac{1}{64} (2 r-s+2) (2 r-s+4) (-2 r+s-3)^2,\qquad  \lambda_{k,l} =\sqrt{2}\, k-\frac{l}{\sqrt{2}},
\ee
and the effective $R_{m,n}^{\text{eff}}$ coefficients are given by:
\be
\label{5.29}
R_{m,n}^{\text{eff}}= \begin{cases}
\frac{4 (-1)^m }{16^{2 m^2}}m^2 \left(16 m^2-1\right), & \text{for  } n= 2 m,\text{  with  }  \,\Delta_{m,2m} = - \frac{1}{8}, \\
-\frac{(-1)^m}{16^{m(2 m+1)}} m (2 m+1) (4 m+1)^2, & \text{for  } n= 2 m +1,  \text{  with  }  \,\Delta_{m,2m+1} = 0, \\
-\frac{(-1)^m}{16^{m(2 m-1)}} m (2 m-1) (4 m-1)^2,   & \text{for  }   n= 2 m -1, \text{  with  }   \,\Delta_{m,2m-1} = 0. \\
0 & \text{otherwise.} 
\end{cases}
\ee
where we wrote explicitly the corresponding value of the pole $\Delta_{m,n}$. 
The result for the $R_{m,n}^{\text{eff}}$ implies now an Ansatz for the recursion as
\be
\label{5.30}
\confblock(H, \Delta, -\frac{3}{32},\frac{21}{32},\frac{45}{32},-\frac{3}{32}, x) =1+\frac{\mathcal{H}_1}{8 \Delta +1}+\frac{\mathcal{H}_2}{8 \Delta }.
\ee
Inserting this Ansatz in the recursion relation, one gets
\be
\label{5.31}
\frac{\mathcal{H}_1}{8 \Delta +1}+\frac{\mathcal{H}_2}{8 \Delta } = \sum_{m,n=1}^{\infty}{\frac{16^{m n} q^{m n} R_{m,n}^{\text{eff}}}{\Delta-\Delta_{m,n}} 
\left(1+\frac{\mathcal{H}_1}{8 (\Delta_{m,n} + m n) +1}+\frac{\mathcal{H}_2}{8 (\Delta_{m,n} + m n) }  \right)}.
\ee
Using the  expressions for the $R_{m,n}^{\text{eff}}$ and separating the contributions coming from the two poles, one gets, after some simple algebra, an equation
for the two functions $\mathcal{H}_1$ and $\mathcal{H}_2$. As in the previous cases this allows to write the $\mathcal{H}_i$ as a combination of series.
The result is slightly more involved than in the one pole cases and reads
\begin{align}
\label{5.32}
\mathcal{H}_1= & \frac{32 \left(\left(\,\mathcal{S}_1+1\right) \,\mathcal{S}_A-8 \,\mathcal{S}_2 \,\mathcal{S}_c\right)}{\left(\,\mathcal{S}_1+1\right) \left(1-2 \,\mathcal{S}_B\right)+256 \,\mathcal{S}_3 \,\mathcal{S}_c},\\
 \mathcal{H}_2= & \frac{8 \left(32 \,\mathcal{S}_3 \,\mathcal{S}_A+\,\mathcal{S}_2 \left(1-2 \,\mathcal{S}_B\right)\right)}{\left(\,\mathcal{S}_1+1\right) \left(2 \,\mathcal{S}_B-1\right)-256 \,\mathcal{S}_3 \,\mathcal{S}_c},\notag
\end{align}
 where the $\mathcal{S}_j$ are the following series
\begin{align}
\label{5.33}
\mathcal{S}_A=&\sum_{m=1}^{\infty}(-1)^m m^2 \left(16 m^2-1\right) q^{2 m^2}, \notag\\
\mathcal{S}_B=&\sum_{m=1}^{\infty}(-1)^m \left(16 m^2-1\right) q^{2 m^2}, \notag \\
\mathcal{S}_C=&\sum_{m=1}^{\infty}(-1)^m m^2 q^{2 m^2}, \\
\mathcal{S}_1=&\sum_{m=1}^{\infty}(-1)^m \left((4 m-1)^2 q^{m (2 m-1)}+(4 m+1)^2 q^{m (2 m+1)}\right), \notag\\
\mathcal{S}_2=&\sum_{m=1}^{\infty}(-1)^m m \left((4 m-1)^2 (2 m-1) q^{m (2 m-1)}+(2 m+1) (4 m+1)^2 
q^{m (2 m+1)}\right), \notag\\
\mathcal{S}_3=&\sum_{m=1}^{\infty}(-1)^m m \left( (2 m-1) q^{m (2 m-1)}+ (2 m+1) q^{m (2 m+1)}\right). \notag
 \end{align}
We do not attempt to put (\ref{5.30}) in the form (\ref{5.27}), but this equivalence may be checked 
at very high order in $q$ by exploiting the explicit sums in (\ref{5.33}). This proves in a very non-trivial 
way the power of the effective recursion approach.

\subsection*{Complete prepotential}

As we did in the previous cases we can collect our results and give the complete prepotential in closed form. 
Combining (\ref{5.3}) and the logarithm of the instanton part (\ref{5.24}) we obtain
\begin{align}
\label{5.34}
F(\bm{\eps}, a, \bm{m}) =&  \,2\, \eps_1^2 \log \frac{a}{\Lambda} + \frac{1}{2} \eps _1^2 \log \left( 1 + \frac{\eps _1^2 \left(-4\, \E_2 +6 \vartheta _2^4+3 \vartheta _4^4\right)}{16\, a^2}  + 
\right.\\ & \qquad\qquad\qquad \left.  \frac{\eps _1^4 \left(4\, \E_2^2-12\, \E_2 \vartheta _2^4+\vartheta _4^4 \left(11 \vartheta _2^4-6\, \E_2\right)+11 \vartheta _2^8+2 \vartheta _4^8\right)}{768\, a^4} \right).\notag
\end{align}  

\section{Systematical study of one and two poles partition functions}
\subsection{The set $\Pi_{1}$ of  one-pole partition functions}
\label{sec:1P}

In the previous sections, we  provided special one-pole partition functions
showing how to extend to the 
$N_{f}=4$ theory the results derived in \cite{Beccaria:2016nnb} for the $\mc N=2^{*}$ theory. In general, 
one has a set $\Pi_{1}$ of special one-pole points $(\bm{\alpha}, \beta)$ where
 $\bm{\alpha}$ and $\beta$
are the following fixed ratios
\be
\label{6.1}
\bm{\alpha} = \bigg( \frac{m_1}{\eps_1}, \frac{m_2}{\eps_1}, 
\frac{m_3}{\eps_1}, \frac{m_4}{\eps_1} \bigg), \qquad
\beta = \frac{\eps_{2}}{\eps_{1}}.
\ee
In terms of the dual CFT, the ratio $\beta$ determines the central charge, while $\bm{\alpha}$
determine the conformal dimensions $\Delta_{1,2,3,4}$ of the relevant 4-point function, according to
(\ref{3.20}).
Due to the symmetry of the partition function under a permutation of $\bm{m}$, it is obvious that a simple permutation 
of the $\alpha_i$ will lead to the same result for $Z\inst$ (but not to the same set of $\Delta_i$ in the dual CFT, see below).
Moreover, one can check that if a set $\bm \alpha$ gives a  one-pole partition function, then
any arbitrary sign choice for the $\alpha_i$ still gives a one-pole $Z\inst$ that may be different from the 
initial one. 
The   position of the pole  is uniquely determined by value of the ratio $\beta$. For instance, 
with these definitions, the $c=1$ case discussed in Sec. (\ref{sec:ramond}) belongs to the group of special points 
\be
\label{6.2}
\bm{\alpha} = \bigg(\frac{1}{2}, \frac{1}{2}, 1, 1 \bigg),\qquad \beta=-1,
\ee 
while the one-pole example in Sec. (\ref{sec:lcft}) is part of the group
\be
\label{6.3}
\bm{\alpha} =  \bigg( \frac{1}{4}, \frac{1}{4}, \frac{1}{2},1 \bigg), \qquad
\beta = -\frac{1}{2}.
\ee 
As stated above, if one is interested in the precise correspondence between a special point and its dual conformal block, the order of the
$\alpha_i$ is important, even if $Z\inst$ is invariant for permutations.
Indeed, the external dimensions for the conformal block in the dual CFT, written as functions of 
$\bm{\alpha}$ and $\beta$ are, see again (\ref{3.20}), 
\begin{align}
\label{6.4}
&\Delta _1 = \frac{\left(\alpha _1-\alpha _2+\beta +1\right) \left(-\alpha _1+\alpha _2+\beta +1\right)}{4 \beta },\notag\\
&\Delta _2 = \frac{\left(-\alpha _1-\alpha _2+\beta +1\right)\left(\alpha _1+\alpha _2+\beta +1\right)}{4 \beta },\\
&\Delta _3 = \frac{\left(-\alpha _3-\alpha _4+\beta +1\right) \left(\alpha _3+\alpha _4+\beta +1\right)}{4 \beta },\notag\\
&\Delta _4 = \frac{\left(\alpha _3-\alpha _4+\beta +1\right) \left(-\alpha _3+\alpha _4+\beta +1\right)}{4 \beta },\notag
\end{align}
and are \emph{not} invariant for permutations of the $\alpha_i$. Looking
again at the  $\Pi_{1}$ point in (\ref{6.2}), we see that external dimensions 
associated to the $\bm{\alpha} = ( \frac{1}{2}, \frac{1}{2}, 1,1 )$ are
$\bm{\Delta} = (0,\frac{1}{4},1,0)$. Instead, the exchange $\alpha_{2}\leftrightarrow \alpha_{3}$
 gives $\bm{\Delta}=(\frac{1}{16}, \frac{9}{16}, \frac{9}{16}, \frac{1}{16})$
that are the values we used in (\ref{4.2}) to exploit the Ramond scalar CFT realization.
The symmetry of $Z\inst$ ensures that the two conformal blocks have the same analytic form, as one can 
explicitly check order by order in the instanton expansion. 

\medskip
The examples (\ref{6.2}) and (\ref{6.3})
are part of a complete list that may be found in various ways, for instance by a brute force
analysis of the Nekrasov functions or, alternatively, by the method suggested in \cite{Nemkov:2016udj}. 
We exploited these approaches, but we also devised a more interesting search algorithm that is 
based on the special simple form of $F\pert$ at the points $\Pi_{N}$. As we observed in the 
illustrative examples and in the analysis of \cite{Beccaria:2016nnb}, the perturbative prepotential
is essentially the sum of $\log(1-\nu_{k}^{2}/\nu^{2})$ over all pole positions $\nu_{k}$. This is a peculiar 
property of the points $\Pi_{N}$ and may be exploited to propose candidate solutions $(\bm{\alpha}, \beta)$.
This method is mandatory beyond the one-pole case and is described in App.~(\ref{sec:algorithm}).
Each  solution in $\Pi_{1}$  has 
input data $(\bm{\alpha}, \beta)$ from which one can write the instanton partition function in the general form 
\be
\label{6.5}
Z ^{\text{inst}}=\left( \frac{x}{16 q}  \right)^{\frac{\nu ^2}{4 \beta }} \left(1-x \right)^{X_1} \vartheta_3(q)^{X_2}\, H_{\nu}(q),
\ee
where, see (\ref{3.23})
\be
\label{6.6}
X_1 = \frac{\left(\beta +1+\sum_{i=1}^4 \alpha_i\right){}^2}{4 \beta },\qquad
X_2 =\frac{-(\beta +1)^2 + 2 \sum_{i=1}^4 \alpha_i^2}{\beta },
\ee
and the function $H_{\nu}(q)$ can be written  (the combination inside the square brackets can be identified with $h_1$, in the large $a$ expansion (\ref{3.6}))
\be
\label{6.7}
H_{\nu}(q) = \frac{\nu ^2+\frac{1}{\beta }\left[ c_1(\bm{\alpha}, \beta) \,\E_2 + c_2(\bm{\alpha}, \beta) \,\vartheta_2^4 + c_3(\bm{\alpha}, \beta) \,\vartheta_4^4     \right]}{(\nu^2-\nu_1^2)},
\ee
where $\nu_1$ is the position of the pole, and with 
\begin{align}
\label{6.8}
c_1(\bm{\alpha}, \beta) &=\frac{1}{24} \left( 2 \,S_{2}-(\beta +1)^2+\beta \right) \left(2\,S_{2}
-(\beta+1)^2+3 \beta \right),\notag\\
c_2(\bm{\alpha}, \beta) &=\frac{1}{12}\,S_{4}-\frac{1}{6}\,S_{2,2}-2\,S_{1,1,1,1},\notag\\
c_3(\bm{\alpha}, \beta) &= \frac{1}{6}\,S_{4}-\frac{1}{3}\,S_{2,2},
\end{align}
and
\be
\label{6.9}
S_{2} = \sum_{i}\alpha_{i}^{2}, \qquad
S_{4} = \sum_{i}\alpha_{i}^{4},\qquad
S_{2,2} = \sum_{i<j}\alpha_{i}^{2}\,\alpha_{j}^{2}, \qquad
S_{1,1,1,1} = \alpha_{1}\,\alpha_{2}\,\alpha_{3}\,\alpha_{4}.
\ee
The quantities in (\ref{6.6}) and (\ref{6.9}) are invariant under permutations. Different choices of signs
of the four $\alpha_{i}$ modify only $X_{1}$ and $S_{1,1,1,1}$, {\em i.e.} the coefficient $c_{2}$.\\

Before listing all the solutions, we complete the collection of formulas writing also the closed form for the perturbative part, which is given by
\begin{align}
\label{6.10}
F\pert = &-\frac{1}{2} \eps_1^2 \left(\beta(\beta+1) -2 S_2+1\right) \log\left( \frac{a}{\Lambda} \right) - 4\,a^2 \log 2 \\ 
&-\beta \eps_1^2 \log \left( 1-\frac{\eps_1^2\, \nu_1^2 }{4 a^2}   \right). \notag
\end{align}
Putting together the above perturbative part with the logarithm of the instanton partition function (\ref{6.5}), one gets the complete prepotential. Note that the arguments 
of the logarithms nicely combine, and the 
resulting formula (discarding as usual all the $a$ independent terms) is very compact
\begin{align}
\label{6.11}
F &= F\pert + F\inst =  -\frac{1}{2} \eps_1^2 \left(\beta(\beta+1) -2 S_2+1\right) \log\left( \frac{a}{\Lambda} \right)  \\
& - \beta\,  \eps_1^2 \log \left(1+ \frac{\eps_1^2}{4\, a^2 \beta }\left( c_1(\bm{\alpha}, \beta) \,\E_2 + c_2(\bm{\alpha}, \beta) \,\vartheta_2^4 + c_3(\bm{\alpha}, \beta) \,\vartheta_4^4     \right)\right).\notag
\end{align}

In Tab.~(\ref{tab:1}) and (\ref{tab:2}), we give the complete list of the groups of one-pole solutions. 
In particular Tab.~(\ref{tab:1}) collects all the special points such that the single pole is already present in the one-instanton
Nekrasov function $Z_1$. For the special points in Tab.~(\ref{tab:2}), $Z_1$ has no poles, and the single pole is present starting from the two-instanton partition function $Z_2$.
All these solutions have been further cross-checked with a high instanton explicit calculation. 
According to \cite{Nemkov:2016udj}, modular transformation properties imply that 
the number of poles of a spherical finite pole block is given by the nice expression
\be
\label{6.12}
\#\ \text{poles} = \left\lfloor \sum_{i=1}^{4}\Delta_{i}-\frac{c+1}{8}\right\rfloor.
\ee
Consistently with our remarks, this expression depends on $\bm \alpha$ through the combination $S_{2}$
 and is thus independent on permutations and change of signs of $\bm{\alpha}$. One checks that the 
 expression in (\ref{6.12}) is indeed equal to 1 for the points $\Pi_{1}$ listed in Tab.~(\ref{tab:1}) and~(\ref{tab:2}). We also remark that the analysis of \cite{Nemkov:2016udj} shows that finite pole
 conformal blocks may only be found at minimal model values of the central charge, 
 $c=1-6\frac{(n-m)^{2}}{n m}$, with $n,m$ being positive coprime integers. This implies that $\beta$ must be a negative rational number. Again, this statement is true in all the solutions we found.

As a final comment, we remind that after equation (\ref{3.8}), we remarked that the $h_k$ 
combinations are quasi-modular forms, taking into account the transformation properties of the $SO(8)$ invariant combinations of masses.
Once the masses are fixed, as in the special  points, this is of course no longer true. 
Nevertheless, from the explicit form of $c_2$, one can note that there are only two possible outcomes for 
the prepotential within each group of  special points. In fact, two distinct outcomes arise only for the special points in Tab.~(\ref{tab:1}).
For the special points in  Tab.~(\ref{tab:2}) the prepotential is unique for each group, since one of the $\alpha_i$ is zero.
It turns out that the $\mathcal{T}$ generator of the modular transformations
maps the two outcomes  into each other for each group in Tab.~(\ref{tab:1}), while the single prepotentials for any group in  Tab.~(\ref{tab:2}) are $\mathcal{T}$-invariant.\\

\begin{table}[h!]
\begin{center}
\renewcommand{\arraystretch}{1.7}
  \begin{tabular}{ c|c|c|cc|ccc}
    \toprule
Special point, $\beta$ and $\bm \alpha$  & CFT $c$ & $\nu_1$ &$X_1$ &  $X_2$ & $c_1$ & $c_2$ & $c_3$ \\ 
\hline 
\rowcolor[gray]{0.925}   
$ \beta = -3 $ , $ \left\{\frac{1}{2},\frac{3}{2},3,1\right\} $ & $ -7 $ & $ 2 $ & $ -\frac{4}{3} $ & $ -7 $ & $ 9 $ & $ -3 $ & $ -3 $ \\
\hline
$ \beta = -2 $ , $ \left\{\frac{1}{2},\frac{1}{2},2,1\right\} $ & $ -2 $ & $ 1 $ & $ -\frac{9}{8} $ & $ -5 $ & $ \frac{4}{3} $ & $ -\frac{2}{3} $ & $ -\frac{2}{3} $ \\
\hline
\rowcolor[gray]{0.925}   
$ \beta = -2 $ , $ \left\{\frac{1}{2},\frac{3}{2},2,1\right\} $ & $ -2 $ & $ 1 $ & $ -2 $ & $ -7 $ & $ 4 $ & $ -4 $ & $ -4 $ \\
\hline
$ \beta = -\frac{3}{2} $ , $ \left\{\frac{3}{4},\frac{1}{4},1,\frac{3}{2}\right\} $ & $ 0 $ & $ \frac{1}{2} $ & $ -\frac{3}{2} $ & $ -5 $ & $ \frac{3}{4} $ & $ -\frac{3}{4} $ & $ -\frac{3}{4} $ \\
\hline
\rowcolor[gray]{0.925}   
$ \beta = -\frac{3}{2} $ , $ \left\{\frac{3}{4},\frac{5}{4},1,\frac{3}{2}\right\} $ & $ 0 $ & $ \frac{1}{2} $ & $ -\frac{8}{3} $ & $ -7 $ & $ \frac{9}{4} $ & $ -\frac{15}{4} $ & $ -\frac{15}{4} $ \\
\hline
$ \beta = -1 $ , $ \left\{\frac{1}{2},1,\frac{1}{2},1\right\} $ & $ 1 $ & $ 0 $ & $ -\frac{9}{4} $ & $ -5 $ & $ \frac{1}{3} $ & $ -\frac{2}{3} $ & $ -\frac{2}{3} $ \\
\hline
\rowcolor[gray]{0.925}   
$ \beta = -\frac{2}{3} $ , $ \left\{\frac{1}{2},\frac{1}{6},1,\frac{2}{3}\right\} $ & $ 0 $ & $ \frac{1}{3} $ & $ -\frac{8}{3} $ & $ -5 $ & $ \frac{4}{27} $ & $ -\frac{4}{27} $ & $ -\frac{4}{27} $ \\
\hline
$ \beta = -\frac{2}{3} $ , $ \left\{\frac{1}{2},\frac{5}{6},1,\frac{2}{3}\right\} $ & $ 0 $ & $ \frac{1}{3} $ & $ -\frac{25}{6} $ & $ -7 $ & $ \frac{4}{9} $ & $ -\frac{20}{27} $ & $ -\frac{20}{27} $ \\
\hline
\rowcolor[gray]{0.925}   
$ \beta = -\frac{1}{2} $ , $ \left\{\frac{1}{4},\frac{1}{4},\frac{1}{2},1\right\} $ & $ -2 $ & $ \frac{1}{2} $ & $ -\frac{25}{8} $ & $ -5 $ & $ \frac{1}{12} $ & $ -\frac{1}{24} $ & $ -\frac{1}{24} $ \\
\hline
$ \beta = -\frac{1}{2} $ , $ \left\{\frac{1}{4},\frac{3}{4},\frac{1}{2},1\right\} $ & $ -2 $ & $ \frac{1}{2} $ & $ -\frac{9}{2} $ & $ -7 $ & $ \frac{1}{4} $ & $ -\frac{1}{4} $ & $ -\frac{1}{4} $ \\
\hline
\rowcolor[gray]{0.925}   
$ \beta = -\frac{1}{3} $ , $ \left\{\frac{1}{2},\frac{1}{6},\frac{1}{3},1\right\} $ & $ -7 $ & $ \frac{2}{3} $ & $ -\frac{16}{3} $ & $ -7 $ & $ \frac{1}{9} $ & $ -\frac{1}{27} $ & $ -\frac{1}{27} $ \\
    \bottomrule
  \end{tabular}
\caption{
First part of the complete list of the one-pole special points $\Pi_{1}=\{(\bm{\alpha}, \beta)\}$.  This first group of solutions corresponds to cases where the pole is already 
present in the first non trivial Nekrasov function $Z_1$. The central charge $c$ of the corresponding CFT is obtained from (\ref{3.20}).
In the other columns we report, as example, the values for the two non-trivial exponents $X_{1}$ and $X_{2}$ and the coefficients $c_{i}$ for the particular choice  $(++++)$ for the signs of $\bm{\alpha}$.
All the other values can be obtained from the formulas in the text. 
}  
  \label{tab:1}
  \end{center}
\end{table}

\begin{table}[h!]
\begin{center}
\renewcommand{\arraystretch}{1.7}
  \begin{tabular}{ c|c|c|cc|ccc}
    \toprule
Special point, $\beta$ and $\bm \alpha$  & CFT $c$ & $\nu_1$ &$X_1$ &  $X_2$ & $c_1$ & $c_2$ & $c_3$ \\ 
\hline 
\rowcolor[gray]{0.925}   
$ \beta =  -4 $  ,  $\left\{0,\frac{3}{2},2,\frac{7}{2}\right\} $  &  $-\frac{25}{2} $  &  $2 $  &  $-1 $  &  $-7 $  &  $16 $  &  $0 $  &  $0 $ \\
\hline 
$ \beta = -3 $  ,  $\left\{0,1,\frac{3}{2},\frac{5}{2}\right\} $  &  $-7 $  &  $1 $  &  $-\frac{3}{4} $  &  $-5 $  &  $3 $  &  $0 $  &  $0 $ \\
\hline 
\rowcolor[gray]{0.925}   
$ \beta = -3 $  ,  $\left\{0,\frac{3}{2},2,\frac{5}{2}\right\} $  &  $-7 $  &  $1 $  &  $-\frac{4}{3} $  &  $-7 $  &  $9 $  &  $-3 $  &  $-3 $ \\
\hline 
$ \beta = -2 $  ,  $\left\{0,\frac{1}{2},1,\frac{5}{2}\right\} $  &  $-2 $  &  $2 $  &  $-\frac{9}{8} $  &  $-7 $  &  $4 $  &  $2 $  &  $2 $ \\
\hline 
\rowcolor[gray]{0.925}   
$ \beta = -2 $  ,  $\left\{0,1,\frac{3}{2},\frac{3}{2}\right\} $  &  $-2 $  &  $0 $  &  $-\frac{9}{8} $  &  $-5 $  &  $\frac{4}{3} $  &  $-\frac{2}{3} $  &  $-\frac{2}{3} $ \\
\hline 
$ \beta = -\frac{3}{2} $  ,  $\left\{0,\frac{1}{4},\frac{1}{2},\frac{9}{4}\right\} $  &  $0 $  &  $2 $  &  $-\frac{25}{24} $  &  $-7 $  &  $\frac{9}{4} $  &  $\frac{15}{8} $  &  $\frac{15}{8} $ \\
\hline 
\rowcolor[gray]{0.925}   
$ \beta = -\frac{3}{2} $  ,  $\left\{0,\frac{1}{2},\frac{3}{4},\frac{7}{4}\right\} $  &  $0 $  &  $1 $  &  $-\frac{25}{24} $  &  $-5 $  &  $\frac{3}{4} $  &  $\frac{3}{8} $  &  $\frac{3}{8} $ \\
\hline 
$ \beta = -\frac{4}{3} $  ,  $\left\{0,\frac{2}{3},\frac{5}{6},\frac{3}{2}\right\} $  &  $\frac{1}{2} $  &  $\frac{2}{3} $  &  $-\frac{4}{3} $  &  $-5 $  &  $\frac{16}{27} $  &  $0 $  &  $0 $ \\
\hline 
\rowcolor[gray]{0.925}   
$ \beta = -1 $  ,  $\left\{0,0,\frac{1}{2},\frac{3}{2}\right\} $  &  $1 $  &  $1 $  &  $-1 $  &  $-5 $  &  $\frac{1}{3} $  &  $\frac{1}{3} $  &  $\frac{1}{3} $ \\
\hline 
$ \beta = -1 $  ,  $\left\{0,\frac{1}{2},1,\frac{3}{2}\right\} $  &  $1 $  &  $1 $  &  $-\frac{9}{4} $  &  $-7 $  &  $1 $  &  $0 $  &  $0 $ \\
\hline 
\rowcolor[gray]{0.925}   
$ \beta = -\frac{3}{4} $  ,  $\left\{0,\frac{1}{2},\frac{5}{8},\frac{9}{8}\right\} $  &  $\frac{1}{2} $  &  $\frac{1}{2} $  &  $-\frac{25}{12} $  &  $-5 $  &  $\frac{3}{16} $  &  $0 $  &  $0 $ \\
\hline 
$ \beta = -\frac{2}{3} $  ,  $\left\{0,\frac{1}{6},\frac{1}{3},\frac{3}{2}\right\} $  &  $0 $  &  $\frac{4}{3} $  &  $-\frac{49}{24} $  &  $-7 $  &  $\frac{4}{9} $  &  $\frac{10}{27} $  &  $\frac{10}{27} $ \\
\hline 
\rowcolor[gray]{0.925}   
$ \beta = -\frac{2}{3} $  ,  $\left\{0,\frac{1}{3},\frac{1}{2},\frac{7}{6}\right\} $  &  $0 $  &  $\frac{2}{3} $  &  $-\frac{49}{24} $  &  $-5 $  &  $\frac{4}{27} $  &  $\frac{2}{27} $  &  $\frac{2}{27} $ \\
\hline 
$ \beta = -\frac{1}{2} $  ,  $\left\{0,\frac{1}{4},\frac{1}{2},\frac{5}{4}\right\} $  &  $-2 $  &  $1 $  &  $-\frac{25}{8} $  &  $-7 $  &  $\frac{1}{4} $  &  $\frac{1}{8} $  &  $\frac{1}{8} $ \\
\hline 
\rowcolor[gray]{0.925}   
$ \beta = -\frac{1}{2} $  ,  $\left\{0,\frac{1}{2},\frac{3}{4},\frac{3}{4}\right\} $  &  $-2 $  &  $0 $  &  $-\frac{25}{8} $  &  $-5 $  &  $\frac{1}{12} $  &  $-\frac{1}{24} $  &  $-\frac{1}{24} $ \\
\hline 
$ \beta = -\frac{1}{3} $  ,  $\left\{0,\frac{1}{3},\frac{1}{2},\frac{5}{6}\right\} $  &  $-7 $  &  $\frac{1}{3} $  &  $-\frac{49}{12} $  &  $-5 $  &  $\frac{1}{27} $  &  $0 $  &  $0 $ \\
\hline 
\rowcolor[gray]{0.925}   
$ \beta = -\frac{1}{3} $  ,  $\left\{0,\frac{1}{2},\frac{2}{3},\frac{5}{6}\right\} $  &  $-7 $  &  $\frac{1}{3} $  &  $-\frac{16}{3} $  &  $-7 $  &  $\frac{1}{9} $  &  $-\frac{1}{27} $  &  $-\frac{1}{27} $ \\
\hline 
$ \beta = -\frac{1}{4} $  ,  $\left\{0,\frac{3}{8},\frac{1}{2},\frac{7}{8}\right\} $  &  $-\frac{25}{2} $  &  $\frac{1}{2} $  &  $-\frac{25}{4} $  &  $-7 $  &  $\frac{1}{16} $  &  $0 $  &  $0 $ \\
    \bottomrule
  \end{tabular}
\caption{
Second part of the complete list of the one-pole special points $\Pi_{1}=\{(\bm{\alpha}, \beta)\}$.  The second group of solutions corresponds to the cases where $Z_1$ has no poles,
and the single pole appears starting from $Z_2$.  The content of the table is as in Tab.~(\ref{tab:1}). }  
  \label{tab:2}
  \end{center}
\end{table}

\subsection{The set $\Pi_{2}$ of two-poles  partition functions}
\label{sec:2P}

Using the algorithm in Sec.~(\ref{sec:algorithm}), we can extend the analysis presented in the previous section and consider  the complete list of 
special points $\Pi_{2}=\{(\bm{\alpha}, \beta)\}$, leading to two poles solutions. Again, we are able to write a closed form for the instanton partition function  for all the $\Pi_{2}$ special points.
Of course, the complete $Z\inst$ is still of the form (\ref{6.5}), so we write here only the function $H_{\nu}(q)$, that for the 
$\Pi_{2}$ cases is given by
\be
\label{6.13}
H_{\nu}(q) = \frac{\nu ^4+ \nu ^2  \frac{1}{\beta} \mathcal{H}_1+\frac{1}{\beta}\left( \frac{1}{2 \beta} \mathcal{H}_1^2 +\mathcal{H}_2\right)}{\left(\nu ^2-\nu _1^2\right) \left(\nu ^2-\nu _2^2\right)}.
\ee
In the previous formula $\nu_{1,2}$ are the positions of the two poles, $\mathcal{H}_1$ is the same combination $c_1(\bm{\alpha}, \beta) \,\E_2 + c_2(\bm{\alpha}, \beta) \,\vartheta_2^4 + c_3(\bm{\alpha}, \beta) \,\vartheta_4^4  $ 
defined in the previous section, see (\ref{6.7}) and (\ref{6.8}), and finally $\mathcal{H}_2$ is given by
\begin{align}
\label{6.14}
\mathcal{H}_2 = & d_1(\bm{\alpha}, \beta) \E_2 \,\vartheta_2^4  +   d_2(\bm{\alpha}, \beta) \,\vartheta_2^8 + d_3(\bm{\alpha}, \beta)   \E_2 \,\vartheta_4^4 +  d_4(\bm{\alpha}, \beta)  \,\vartheta_4^8  + \\  
& d_5(\bm{\alpha}, \beta) \,\vartheta_2^4  \,\vartheta_4^4 + d_6(\bm{\alpha}, \beta) \E_2^2 + 
d_7(\bm{\alpha}, \beta)\, \E_4  \notag.
\end{align}
The seven coefficients $d_i(\bm{\alpha}, \beta), i=1,\dots, 7$, are
\begin{align}
\label{6.15}
d_1(\bm{\alpha}, \beta)  = & \frac{1}{36} \left(-2 S_{2,2}-24 S_{1,1,1,1}+S_4\right) \left(2 S_2-(\beta -1)^2\right), \notag\\
d_2(\bm{\alpha}, \beta)  = & -\frac{1}{36} \left(-2 S_{2,2}-24 S_{1,1,1,1}+S_4\right) \left(S_2-2 (\beta  (\beta +1)+1)\right), \notag\\
d_3(\bm{\alpha}, \beta)  = & \frac{1}{18} \left(S_4-2 S_{2,2}\right) \left(2 S_2-(\beta -1)^2\right), \notag\\
d_4(\bm{\alpha}, \beta)  = & \frac{1}{18} \left(S_4-2 S_{2,2}\right) \left(S_2-2 (\beta  (\beta +1)+1)\right), \\
d_5(\bm{\alpha}, \beta)  = & \frac{1}{18} \left(-2 S_{2,2}+24 S_{1,1,1,1}+S_4\right) 
\left(S_2-2 (\beta  (\beta +1)+1)\right),\notag \\
d_6(\bm{\alpha}, \beta)  = & -\frac{1}{144} \left((\beta -1)^2-2 S_2\right) \left((\beta -1) \beta -2 S_2+1\right) \left(\beta  (\beta +1)-2 S_2+1\right), \notag\\
d_7(\bm{\alpha}, \beta)  = & \frac{1}{720} \left(-24 S_{2,4}+432 S_{2,2,2}-13 \beta ^6-10 \beta ^5+\beta ^4 \left(54 S_2-17\right)+\beta ^3 \left(40 S_2-10\right)  \right.\notag\\   
&\left. +\beta ^2 \left(-60 S_2^2+70 S_2-17\right)+\beta  \left(-40 S_2^2+40 S_2-10\right)+8 S_2^3-
\right.\notag\\   
&\left. 60 S_2^2+54 S_2+24 S_6-13\right), \notag
\end{align}
where we introduced the additional quantities (extending the list in (\ref{6.9}))
\be
\label{6.16}
S_{6} = \sum_{i}\alpha_{i}^{6},\qquad
S_{2,4} = \sum_{i\neq j}\alpha_{i}^{2}\,\alpha_{j}^{4}, \qquad
S_{2,2,2} = \sum_{i < j < k}\alpha_{i}^{2}\,\alpha_{j}^{2}\,\alpha_{k}^{2}.
\ee
In all cases, the perturbative prepotential takes the following natural  closed form 
\begin{align}
\label{6.17}
F\pert = &-\frac{1}{2} \eps_1^2 \left(\beta(\beta+1) -2 S_2+1\right) \log\left( \frac{a}{\Lambda} \right) - 4\,a^2 \log 2 \\ 
&-\beta\, \eps_1^2 \log \left( \frac{\,\eps _1^4}{16\, a^4}\,\left(\frac{4 a^2}{\eps _1^2}-\nu _1^2\right) \left(\frac{4 a^2}{\eps_1^2}-\nu _2^2\right) \right). \notag
\end{align}
As in the case of 1-pole points, the two logarithms coming from $F\pert $ and 
$F\inst=- \eps_1 \eps_2 \log Z\inst $ combine nicely  in order to 
give the following simplified form of the full prepotential 
\begin{align}
\label{6.18}
F = F\pert + F\inst = &  -\frac{1}{2} \eps_1^2 \left(\beta(\beta+1) -2 S_2+1\right) \log\left( \frac{a}{\Lambda} \right) \notag \\
& - \beta\,  \eps_1^2 \log \left(
1+ \frac{8\, a^2\,  \eps_1^2\,  \beta\,  \mathcal{H}_1 +\eps_1^4 \left(2\, \beta\,  \mathcal{H}_2+\mathcal{H}_1^2\right)}{32\, a^4\, \beta ^2}  \right).
\end{align}
The number of solutions $\Pi_{2}$ is 74, significantly larger than for $\Pi_{1}$, so the result are collected in two tables, Tab.~(\ref{tabb2.1}) and Tab.~(\ref{tabb2.2}).
All the $\Pi_{2}$ points have been tested successfully by comparing them with 
a direct calculation of $Z\inst$ at high instanton number.
One finally checks that the quantity defined in (\ref{6.12}) is equal to 2 in all cases.

\begin{table}[h!]
\begin{center}
\renewcommand{\arraystretch}{1.7}
  \begin{tabular}{ c|c|cc|c|c|c|cc}
    \toprule
Special point, $\beta$ and $\bm \alpha$  & CFT $c$ & $\nu_1$ & $\nu_2$ &  & Special point, $\beta$ and $\bm \alpha$  &  CFT $c$ & $\nu_1$ & $\nu_2$  \\
\hline
\rowcolor[gray]{0.925}   
$ \beta = -6 $   ,   $  \left\{0,\frac{5}{2},3,\frac{11}{2}\right\} $   &   $  -24 $   &   $  2 $   &   $  4 $   &   $  \text{  } $   &   $  \beta =   -1 $   ,   $  \left\{\frac{1}{2},\frac{1}{2},1,2\right\} $   &   $  1 $   &   $  0 $   &   $  2 $ \\
\hline    
$ \beta = -5 $   ,   $  \left\{0,2,\frac{5}{2},\frac{9}{2}\right\} $   &   $  -\frac{91}{5} $   &   $  1 $   &   $  3 $   &   $  \text{  } $   &   $  \beta =  -1 $   ,   $  \left\{0,1,\frac{3}{2},\frac{3}{2}\right\} $   &   $  1 $   &   $  1 $   &   $  1 $ \\
\hline
\rowcolor[gray]{0.925}    
$ \beta = -5 $   ,   $  \left\{0,\frac{5}{2},3,\frac{9}{2}\right\} $   &   $  -\frac{91}{5} $   &   $  1 $   &   $  3 $   &   $  \text{  } $   &   $  \beta =  -\frac{4}{5} $   ,   $  \left\{0,\frac{2}{5},\frac{11}{10},\frac{3}{2}\right\} $   &   $  \frac{7}{10} $   &   $  \frac{2}{5} $   &   $  \frac{6}{5} $ \\
\hline    
$ \beta = -5 $   ,   $  \left\{\frac{1}{2},2,\frac{5}{2},5\right\} $   &   $  -\frac{91}{5} $   &   $  2 $   &   $  4 $   &   $  \text{  } $   &   $  \beta =  -\frac{3}{4} $   ,   $  \left\{\frac{1}{8},\frac{3}{8},1,\frac{3}{2}\right\} $   &   $  \frac{1}{2} $   &   $  \frac{1}{4} $   &   $  \frac{5}{4} $ \\
\hline 
\rowcolor[gray]{0.925}
$ \beta = -4 $   ,   $  \left\{0,2,\frac{5}{2},\frac{7}{2}\right\} $   &   $  -\frac{25}{2} $   &   $  0 $   &   $  2 $   &   $  \text{  } $   &   $  \beta =  -\frac{3}{4} $   ,   $  \left\{\frac{3}{8},\frac{7}{8},1,\frac{3}{2}\right\} $   &   $  \frac{1}{2} $   &   $  \frac{1}{4} $   &   $  \frac{5}{4} $ \\
\hline 
$ \beta = -4 $   ,   $  \left\{\frac{1}{2},\frac{3}{2},2,4\right\} $   &   $  -\frac{25}{2} $   &   $  1 $   &   $  3 $   &   $  \text{  } $   &   $  \beta =  -\frac{3}{4} $   ,   $  \left\{0,\frac{1}{2},\frac{9}{8},\frac{13}{8}\right\} $   &   $  \frac{1}{2} $   &   $  \frac{1}{2} $   &   $  \frac{3}{2} $ \\
\hline 
\rowcolor[gray]{0.925}
$ \beta = -4 $   ,   $  \left\{\frac{1}{2},2,\frac{5}{2},4\right\} $   &   $  -\frac{25}{2} $   &   $  1 $   &   $  3 $   &   $  \text{  } $   &   $  \beta =  -\frac{3}{4} $   ,   $  \left\{0,\frac{1}{2},\frac{5}{8},\frac{15}{8}\right\} $   &   $  \frac{1}{2} $   &   $  \frac{1}{2} $   &   $  2 $ \\
\hline 
$ \beta = -3 $   ,   $  \left\{\frac{1}{2},\frac{3}{2},2,3\right\} $   &   $  -7 $   &   $  0 $   &   $  2 $   &   $  \text{  } $   &   $  \beta =  -\frac{2}{3} $   ,   $  \left\{0,\frac{1}{3},\frac{5}{6},\frac{3}{2}\right\} $   &   $  0 $   &   $  0 $   &   $  \frac{4}{3} $ \\
\hline 
\rowcolor[gray]{0.925}
$ \beta = -3 $   ,   $  \left\{0,1,\frac{3}{2},\frac{7}{2}\right\} $   &   $  -7 $   &   $  1 $   &   $  3 $   &   $  \text{  } $   &   $  \beta =  -\frac{2}{3} $   ,   $  \left\{\frac{1}{2},\frac{2}{3},1,\frac{7}{6}\right\} $   &   $  0 $   &   $  \frac{1}{3} $   &   $  \frac{2}{3} $ \\
\hline 
$ \beta = -3 $   ,   $  \left\{1,\frac{3}{2},\frac{5}{2},3\right\} $   &   $  -7 $   &   $  1 $   &   $  2 $   &   $  \text{  } $   &   $  \beta =  -\frac{2}{3} $   ,   $  \left\{\frac{1}{6},\frac{1}{2},1,\frac{4}{3}\right\} $   &   $  0 $   &   $  \frac{1}{3} $   &   $  1 $ \\
\hline 
\rowcolor[gray]{0.925}
$ \beta = -3 $   ,   $  \left\{0,\frac{3}{2},2,\frac{7}{2}\right\} $   &   $  -7 $   &   $  1 $   &   $  3 $   &   $  \text{  } $   &   $  \beta =  -\frac{2}{3} $   ,   $  \left\{0,\frac{1}{6},\frac{1}{2},\frac{5}{3}\right\} $   &   $  0 $   &   $  \frac{1}{3} $   &   $  \frac{5}{3} $ \\
\hline 
$ \beta = -\frac{5}{2} $   ,   $  \left\{\frac{3}{4},\frac{5}{4},2,\frac{5}{2}\right\} $   &   $  -\frac{22}{5} $   &   $  \frac{1}{2} $   &   $  \frac{3}{2} $   &   $  \text{  } $   &   $  \beta =  -\frac{2}{3} $   ,   $  \left\{\frac{1}{2},\frac{5}{6},1,\frac{4}{3}\right\} $   &   $  0 $   &   $  \frac{1}{3} $   &   $  1 $ \\
\hline 
\rowcolor[gray]{0.925}
$ \beta = -\frac{5}{2} $   ,   $  \left\{\frac{5}{4},\frac{7}{4},2,\frac{5}{2}\right\} $   &   $  -\frac{22}{5} $   &   $  \frac{1}{2} $   &   $  \frac{3}{2} $   &   $  \text{  } $   &   $  \beta =  -\frac{2}{3} $   ,   $  \left\{\frac{1}{6},\frac{2}{3},1,\frac{3}{2}\right\} $   &   $  0 $   &   $  \frac{1}{3} $   &   $  \frac{4}{3} $ \\
\hline 
$ \beta = -\frac{5}{2} $   ,   $  \left\{0,\frac{1}{2},\frac{3}{4},\frac{15}{4}\right\} $   &   $  -\frac{22}{5} $   &   $  2 $   &   $  4 $   &   $  \text{  } $   &   $  \beta =  -\frac{2}{3} $   ,   $  \left\{0,\frac{1}{2},\frac{5}{6},\frac{5}{3}\right\} $   &   $  0 $   &   $  \frac{1}{3} $   &   $  \frac{5}{3} $ \\
\hline 
\rowcolor[gray]{0.925}
$ \beta = -2 $   ,   $  \left\{1,\frac{3}{2},\frac{3}{2},2\right\} $   &   $  -2 $   &   $  0 $   &   $  1 $   &   $  \text{  } $   &   $  \beta =  -\frac{2}{3} $   ,   $  \left\{0,\frac{1}{3},\frac{7}{6},\frac{3}{2}\right\} $   &   $  0 $   &   $  \frac{2}{3} $   &   $  \frac{4}{3} $ \\
\hline 
$ \beta = -2 $   ,   $  \left\{0,1,\frac{3}{2},\frac{5}{2}\right\} $   &   $  -2 $   &   $  0 $   &   $  2 $   &   $  \text{  } $   &   $  \beta =  -\frac{3}{5} $   ,   $  \left\{0,\frac{1}{5},\frac{7}{10},\frac{3}{2}\right\} $   &   $  -\frac{3}{5} $   &   $  \frac{1}{5} $   &   $  \frac{7}{5} $ \\
\hline 
\rowcolor[gray]{0.925}
$ \beta = -2 $   ,   $  \left\{0,\frac{1}{2},\frac{1}{2},3\right\} $   &   $  -2 $   &   $  1 $   &   $  3 $   &   $  \text{  } $   &   $  \beta =  -\frac{3}{5} $   ,   $  \left\{\frac{3}{10},\frac{1}{2},1,\frac{6}{5}\right\} $   &   $  -\frac{3}{5} $   &   $  \frac{2}{5} $   &   $  \frac{4}{5} $ \\
\hline 
$ \beta = -2 $   ,   $  \left\{\frac{1}{2},1,2,\frac{5}{2}\right\} $   &   $  -2 $   &   $  1 $   &   $  2 $   &   $  \text{  } $   &   $  \beta =  -\frac{3}{5} $   ,   $  \left\{0,\frac{7}{10},\frac{4}{5},\frac{3}{2}\right\} $   &   $  -\frac{3}{5} $   &   $  \frac{1}{5} $   &   $  \frac{7}{5} $ \\
\hline 
\rowcolor[gray]{0.925}
$ \beta = -2 $   ,   $  \left\{0,\frac{1}{2},\frac{3}{2},3\right\} $   &   $  -2 $   &   $  1 $   &   $  3 $   &   $  \text{  } $   &   $  \beta =  -\frac{1}{2} $   ,   $  \left\{\frac{1}{2},\frac{3}{4},\frac{3}{4},1\right\} $   &   $  -2 $   &   $  0 $   &   $  \frac{1}{2} $ \\
\hline 
$ \beta = -\frac{5}{3} $   ,   $  \left\{0,\frac{1}{3},\frac{7}{6},\frac{5}{2}\right\} $   &   $  -\frac{3}{5} $   &   $  \frac{1}{3} $   &   $  \frac{7}{3} $   &   $  \text{  } $   &   $  \beta =  -\frac{1}{2} $   ,   $  \left\{0,\frac{1}{2},\frac{3}{4},\frac{5}{4}\right\} $   &   $  -2 $   &   $  0 $   &   $  1 $ \\
    \bottomrule
  \end{tabular}
\caption{First part of the complete list of the two-poles special points. We list the groups $\Pi_{2}=\{(\bm{\alpha},\beta)\}$, the corresponding value of the CFT central charge and the position of the two poles.}  
  \label{tabb2.1}
  \end{center}
\end{table}

\begin{table}[h!]
\begin{center}
\renewcommand{\arraystretch}{1.7}
  \begin{tabular}{ c|c|cc|c|c|c|cc}
    \toprule
Special point, $\beta$ and $\bm \alpha$  & CFT $c$ & $\nu_1$ & $\nu_2$ &  & Special point, $\beta$ and $\bm \alpha$  &  CFT $c$ & $\nu_1$ & $\nu_2$  \\
\hline 
\rowcolor[gray]{0.925}
$ \beta = -\frac{5}{3} $   ,   $  \left\{\frac{1}{2},\frac{5}{6},\frac{5}{3},2\right\} $   &   $  -\frac{3}{5} $   &   $  \frac{2}{3} $   &   $  \frac{4}{3} $   &   $  \text{  } $   &   $  \beta =  -\frac{1}{2} $   ,   $  \left\{0,\frac{1}{4},\frac{1}{4},\frac{3}{2}\right\} $   &   $  -2 $   &   $  \frac{1}{2} $   &   $  \frac{3}{2} $ \\
\hline 
$ \beta = -\frac{5}{3} $   ,   $  \left\{0,\frac{7}{6},\frac{4}{3},\frac{5}{2}\right\} $   &   $  -\frac{3}{5} $   &   $  \frac{1}{3} $   &   $  \frac{7}{3} $   &   $  \text{  } $   &   $  \beta =  -\frac{1}{2} $   ,   $  \left\{\frac{1}{4},\frac{1}{2},1,\frac{5}{4}\right\} $   &   $  -2 $   &   $  \frac{1}{2} $   &   $  1 $ \\
\hline 
\rowcolor[gray]{0.925}
$ \beta = -\frac{3}{2} $   ,   $  \left\{0,\frac{1}{2},\frac{5}{4},\frac{9}{4}\right\} $   &   $  0 $   &   $  0 $   &   $  2 $   &   $  \text{  } $   &   $  \beta =  -\frac{1}{2} $   ,   $  \left\{0,\frac{1}{4},\frac{3}{4},\frac{3}{2}\right\} $   &   $  -2 $   &   $  \frac{1}{2} $   &   $  \frac{3}{2} $ \\
\hline 
$ \beta = -\frac{3}{2} $   ,   $  \left\{\frac{3}{4},1,\frac{3}{2},\frac{7}{4}\right\} $   &   $  0 $   &   $  \frac{1}{2} $   &   $  1 $   &   $  \text{  } $   &   $  \beta =  -\frac{2}{5} $   ,   $  \left\{\frac{3}{10},\frac{1}{2},\frac{4}{5},1\right\} $   &   $  -\frac{22}{5} $   &   $  \frac{1}{5} $   &   $  \frac{3}{5} $ \\
\hline 
\rowcolor[gray]{0.925}
$ \beta = -\frac{3}{2} $   ,   $  \left\{\frac{1}{4},\frac{3}{4},\frac{3}{2},2\right\} $   &   $  0 $   &   $  \frac{1}{2} $   &   $  \frac{3}{2} $   &   $  \text{  } $   &   $  \beta =  -\frac{2}{5} $   ,   $  \left\{\frac{1}{2},\frac{7}{10},\frac{4}{5},1\right\} $   &   $  -\frac{22}{5} $   &   $  \frac{1}{5} $   &   $  \frac{3}{5} $ \\
\hline 
$ \beta = -\frac{3}{2} $   ,   $  \left\{0,\frac{1}{4},\frac{3}{4},\frac{5}{2}\right\} $   &   $  0 $   &   $  \frac{1}{2} $   &   $  \frac{5}{2} $   &   $  \text{  } $   &   $  \beta =  -\frac{2}{5} $   ,   $  \left\{0,\frac{1}{5},\frac{3}{10},\frac{3}{2}\right\} $   &   $  -\frac{22}{5} $   &   $  \frac{4}{5} $   &   $  \frac{8}{5} $ \\
\hline 
\rowcolor[gray]{0.925}
$ \beta = -\frac{3}{2} $   ,   $  \left\{\frac{3}{4},\frac{5}{4},\frac{3}{2},2\right\} $   &   $  0 $   &   $  \frac{1}{2} $   &   $  \frac{3}{2} $   &   $  \text{  } $   &   $  \beta =  -\frac{1}{3} $   ,   $  \left\{\frac{1}{6},\frac{1}{2},\frac{2}{3},1\right\} $   &   $  -7 $   &   $  0 $   &   $  \frac{2}{3} $ \\
\hline 
$ \beta = -\frac{3}{2} $   ,   $  \left\{\frac{1}{4},1,\frac{3}{2},\frac{9}{4}\right\} $   &   $  0 $   &   $  \frac{1}{2} $   &   $  2 $   &   $  \text{  } $   &   $  \beta =  -\frac{1}{3} $   ,   $  \left\{0,\frac{1}{3},\frac{1}{2},\frac{7}{6}\right\} $   &   $  -7 $   &   $  \frac{1}{3} $   &   $  1 $ \\
\hline 
\rowcolor[gray]{0.925}
$ \beta = -\frac{3}{2} $   ,   $  \left\{0,\frac{3}{4},\frac{5}{4},\frac{5}{2}\right\} $   &   $  0 $   &   $  \frac{1}{2} $   &   $  \frac{5}{2} $   &   $  \text{  } $   &   $  \beta =  -\frac{1}{3} $   ,   $  \left\{\frac{1}{3},\frac{1}{2},\frac{5}{6},1\right\} $   &   $  -7 $   &   $  \frac{1}{3} $   &   $  \frac{2}{3} $ \\
\hline 
$ \beta = -\frac{3}{2} $   ,   $  \left\{0,\frac{1}{2},\frac{7}{4},\frac{9}{4}\right\} $   &   $  0 $   &   $  1 $   &   $  2 $   &   $  \text{  } $   &   $  \beta =  -\frac{1}{3} $   ,   $  \left\{0,\frac{1}{2},\frac{2}{3},\frac{7}{6}\right\} $   &   $  -7 $   &   $  \frac{1}{3} $   &   $  1 $ \\
\hline 
\rowcolor[gray]{0.925}
$ \beta = -\frac{4}{3} $   ,   $  \left\{\frac{1}{6},\frac{1}{2},\frac{4}{3},2\right\} $   &   $  \frac{1}{2} $   &   $  \frac{1}{3} $   &   $  \frac{5}{3} $   &   $  \text{  } $   &   $  \beta =  -\frac{1}{4} $   ,   $  \left\{0,\frac{1}{2},\frac{5}{8},\frac{7}{8}\right\} $   &   $  -\frac{25}{2} $   &   $  0 $   &   $  \frac{1}{2} $ \\
\hline 
$ \beta = -\frac{4}{3} $   ,   $  \left\{\frac{1}{2},\frac{7}{6},\frac{4}{3},2\right\} $   &   $  \frac{1}{2} $   &   $  \frac{1}{3} $   &   $  \frac{5}{3} $   &   $  \text{  } $   &   $  \beta =  -\frac{1}{4} $   ,   $  \left\{\frac{1}{8},\frac{3}{8},\frac{1}{2},1\right\} $   &   $  -\frac{25}{2} $   &   $  \frac{1}{4} $   &   $  \frac{3}{4} $ \\
\hline 
\rowcolor[gray]{0.925}
$ \beta = -\frac{4}{3} $   ,   $  \left\{0,\frac{2}{3},\frac{3}{2},\frac{13}{6}\right\} $   &   $  \frac{1}{2} $   &   $  \frac{2}{3} $   &   $  2 $   &   $  \text{  } $   &   $  \beta =  -\frac{1}{4} $   ,   $  \left\{\frac{1}{8},\frac{1}{2},\frac{5}{8},1\right\} $   &   $  -\frac{25}{2} $   &   $  \frac{1}{4} $   &   $  \frac{3}{4} $ \\
\hline 
$ \beta = -\frac{4}{3} $   ,   $  \left\{0,\frac{2}{3},\frac{5}{6},\frac{5}{2}\right\} $   &   $  \frac{1}{2} $   &   $  \frac{2}{3} $   &   $  \frac{8}{3} $   &   $  \text{  } $   &   $  \beta =  -\frac{1}{5} $   ,   $  \left\{0,\frac{2}{5},\frac{1}{2},\frac{9}{10}\right\} $   &   $  -\frac{91}{5} $   &   $  \frac{1}{5} $   &   $  \frac{3}{5} $ \\
\hline 
\rowcolor[gray]{0.925}
$ \beta = -\frac{5}{4} $   ,   $  \left\{0,\frac{1}{2},\frac{11}{8},\frac{15}{8}\right\} $   &   $  \frac{7}{10} $   &   $  \frac{1}{2} $   &   $  \frac{3}{2} $   &   $  \text{  } $   &   $  \beta =  -\frac{1}{5} $   ,   $  \left\{0,\frac{1}{2},\frac{3}{5},\frac{9}{10}\right\} $   &   $  -\frac{91}{5} $   &   $  \frac{1}{5} $   &   $  \frac{3}{5} $ \\
\hline 
$ \beta = -1 $   ,   $  \left\{\frac{1}{2},1,1,\frac{3}{2}\right\} $   &   $  1 $   &   $  0 $   &   $  1 $   &   $  \text{  } $   &   $  \beta =  -\frac{1}{5} $   ,   $  \left\{\frac{1}{10},\frac{2}{5},\frac{1}{2},1\right\} $   &   $  -\frac{91}{5} $   &   $  \frac{2}{5} $   &   $  \frac{4}{5} $ \\
\hline 
\rowcolor[gray]{0.925}
$ \beta = -1 $   ,   $  \left\{0,0,\frac{3}{2},\frac{3}{2}\right\} $   &   $  1 $   &   $  1 $   &   $  1 $   &   $  \text{  } $   &   $  \beta =  -\frac{1}{6} $   ,   $  \left\{0,\frac{5}{12},\frac{1}{2},\frac{11}{12}\right\} $   &   $  -24 $   &   $  \frac{1}{3} $   &   $  \frac{2}{3} $ \\
    \bottomrule
  \end{tabular}
\caption{Second part of the complete list of the two-poles special points. We list the groups $\Pi_{2}=\{(\bm{\alpha},\beta)\}$, the corresponding value of the CFT central charge and the position of the two poles.}  
  \label{tabb2.2}
  \end{center}
\end{table}

%
%

\section*{Acknowledgments}
We  thank Hubert Saleur, Susanne Reffert, and Boris Konopelchenko for comments and clarifying discussions. 

\appendix

\section{Special functions and useful identities}
\label{sec:ids}

In this paper the relation between the modular parameters $\tau$ and $q$ is  $q=e^{i\,\pi\,\tau}$. Accordingly, 
the Dedekind $\eta$ function reads
\be
\label{A.1}
\eta(\tau)=q^{\frac{1}{12}} \prod _{k=1}^\infty (1-q^{2k}).
\ee
The definition of the Jacobi $\vartheta_{2,3,4}(q)$ functions as series expansions around $q=0$ is 
\be
\la{A.2}
\vartheta_{2}(q) = 2\,q^{1/4}\,\sum_{n=0}^{\infty}q^{n(n+1)}, \quad
\vartheta_{3}(q) =1+ 2\,\sum_{n=1}^{\infty}q^{n^{2}}, \quad
\vartheta_{4}(q) =1+ 2\,\sum_{n=1}^{\infty}(-1)^{n}\,q^{n^{2}}.
\ee
The first 3 Eisenstein series are
\be
\la{A.3}
\E_{2}(q) = 1-24\,\sum_{n=1}^{\infty}\frac{n\,q^{n}}{1-q^{n}}, \quad
\E_{4}(q) = 1+240\,\sum_{n=1}^{\infty}\frac{n^{3}\,q^{n}}{1-q^{n}}, \quad
\E_{6}(q) = 1-504\,\sum_{n=1}^{\infty}\frac{n^{5}\,q^{n}}{1-q^{n}}.
\ee
The Eisenstein series can be written in terms of Jacobi $\vartheta$ functions 
(all functions with argument $q$ if not specified) by means of the relations
\be
\la{A.4}
\begin{split}
\E_{2} &= 12\,q\,\frac{d}{dq}\log\vartheta_{3}+\vartheta_{4}^{4}-\vartheta_{2}^{4}, \\
\E_{4} &= \frac{1}{2}\,\left(\vartheta_{2}^{8}+\vartheta_{3}^{8}+\vartheta_{4}^{8}\right) = \left(\vartheta_{2}^{8}+\vartheta_{2}^{4}\vartheta_{4}^{4}+\vartheta_{4}^{8}\right), \\
\E_{6} &= \frac{1}{2}\,
\left(\vartheta_{2}^{4}+\vartheta_{3}^{4}\right)\,
\left(\vartheta_{4}^{4}-\vartheta_{2}^{4}\right)\,
\left(\vartheta_{3}^{4}+\vartheta_{4}^{4}\right).
\end{split}
\ee
If we define, see (\ref{1.1}),
\be
\label{A.5}
x = \frac{\vartheta_{2}^{4}(q)}{\vartheta_{3}^{4}(q)},
\ee
we can write the following useful alternative forms
\be
\la{A.6}
\begin{split}
\E_{2}(q) &= \frac{6}{\pi}\, \mathbb{E}(x)\,\vartheta_{3}^{2}(q)-\vartheta_{3}^{4}(q)-\vartheta_{4}^{4}(q),\\
\E_{4}(q) &= \left[\frac{2\,\mathbb{K}(x)}{\pi}\right]^{4}\,\left[1-x(1-x)\right], \\
\E_{6}(q) &= \left[\frac{2\,\mathbb{K}(x)}{\pi}\right]^{6}\,(1-2x)\,\left[1+\frac{1}{2}x(1-x)\right],
\end{split}
\ee
and also the following identities useful to discuss generic 1-pole partition functions
\be
\la{A.7}
\begin{split}
& k_{1}\,\E_{2}(q)+k_{2}\,\vartheta_{4}^{4}(q)+k_{3}\,\vartheta_{2}^{4}(q) = \\
&\qquad\qquad \frac{4}{\pi^{2}}\,\mathbb{K}(x)\,\bigg\{
3\,k_{1}\,\mathbb{E}(x)+[-2\,k_{1}+k_{2}+(k_{1}-k_{2}+k_{3})\,x]\, \mathbb{K}(x)\bigg\}
\end{split}
\ee
\be
\la{A.8}
\begin{split}
\vartheta^{2}_{2}(q) = \frac{2}{\pi}\,\sqrt{x}\,\mathbb{K}(x), \quad
\vartheta^{2}_{3}(q) = \frac{2}{\pi}\,\mathbb{K}(x), \quad
\vartheta^{2}_{4}(q) = \frac{2}{\pi}\,\sqrt{1-x}\,\mathbb{K}(x).
\end{split}
\ee
Additional useful identities are the well known \emph{aequatio identica satis abstrusa}
\be
\la{A.9}
\vartheta_{3}^{4}-\vartheta_{2}^{4}-\vartheta_{4}^{4}=0,
\ee
and the $q$-derivatives of theta functions
\be
\la{A.10}
\begin{split}
q\,\frac{d}{dq}\log\vartheta_{2}& = \frac{1}{12}\,(\E_{2}+\vartheta_{3}^{4}+\vartheta_{4}^{4}),\\
q\,\frac{d}{dq}\log\vartheta_{3} &= \frac{1}{12}\,(\E_{2}+\vartheta_{2}^{4}-\vartheta_{4}^{4}),\\
q\,\frac{d}{dq}\log\vartheta_{4} &= \frac{1}{12}\,(\E_{2}-\vartheta_{2}^{4}-\vartheta_{3}^{4}).
\end{split}
\ee
We notice also the remarkable duplication identity that has been exploited in the text
\be
\la{A.11}
\vartheta_{2}(\sqrt{q})^{2} = 2\,\vartheta_{2}(q)\,\vartheta_{3}(q).
\ee
This can be proved by using the Eulerian product form of the theta functions. To this aim, 
we define 
\be
\la{A.12}
(n) = \prod_{k=1}^{\infty}(1-q^{k\,n}),
\ee
and remind that   \cite{zucker1990further}
\be
\la{A.13}
\vartheta_{2}(q) = 2\,q^{1/4}\,\frac{(4)^{2}}{(2)}, \quad
\vartheta_{2}(\sqrt{q}) = 2\,q^{1/8}\,\frac{(2)^{2}}{(1)}, \quad
\vartheta_{3}(q) = \frac{(2)^{5}}{(1)^{2}(4)^{2}}.
\ee
From (\ref{A.13}), we readily check (\ref{A.11}).

\section{An efficient algorithm to find the sets $\Pi_{N}$}
\label{sec:algorithm}

Let us illustrate an efficient algorithm that is able to locate special $N$-pole solutions with moderate
computational effort. We examine in details the case of 1-pole solutions. Our starting remark is that 
the perturbative prepotential takes the special form 
\be
\la{B.1}
F\pert = -\beta\,\eps_{1}^{2}\,\log\bigg(1-\eps_{1}^{2}\,\frac{\nu_{1}^{2}}{\nu^{2}}\bigg),
\ee
at all special points. This is an extremely useful Ansatz to locate special points that can be ultimately cross checked by a dedicated high-instanton ( $\ge 20$ )
computation. As a first step, we match the large $a$ expansion of the 
perturbative contribution (\ref{3.4}) with the Ansatz in (\ref{B.1}).
The comparison allows to express the sums $Q_{n} = \sum_{i=1}^{4} \alpha_{i}^{2n}$ with 
$n\ge 2$ in terms of the three unknown quantities $Q_{1}$ , $\nu_{1}^{2}$, and $\beta$. For instance, 
we find 
\begin{align}
\label{B.2}
Q_{2} &= -\frac{\beta ^4}{8}-3 \beta  \nu _1^2+\beta ^2
   \left(\frac{Q_1}{2}-\frac{1}{8}\right)+\frac{Q_1}{2}-\frac{1}{8}, \\
Q_{3} &= -\frac{\beta ^6}{16}-\frac{15}{4} \beta ^3 \nu _1^2+\beta  \left(-\frac{15 \nu
   _1^4}{8}-\frac{15 \nu _1^2}{4}\right)+\beta ^4 \left(\frac{3
   Q_1}{16}-\frac{11}{64}\right)\notag \\
   &+\beta ^2 \left(\frac{5
   Q_1}{8}-\frac{11}{64}\right)+\frac{3 Q_1}{16}-\frac{1}{16},\notag  \\
Q_{4} &= -\frac{3 \beta ^8}{128}-\frac{21}{8} \beta ^5 \nu _1^2+\beta ^3 \left(-\frac{35 \nu
   _1^4}{8}-\frac{35 \nu _1^2}{4}\right)+\beta  \left(-\frac{7 \nu _1^6}{8}-\frac{35 \nu
   _1^4}{8}-\frac{21 \nu _1^2}{8}\right)\notag \\
   &+\beta ^6
   \left(\frac{Q_1}{16}-\frac{9}{64}\right)+\beta ^4 \left(\frac{7
   Q_1}{16}-\frac{21}{128}\right)+\beta ^2 \left(\frac{7
   Q_1}{16}-\frac{9}{64}\right)+\frac{Q_1}{16}-\frac{3}{128},\notag
\end{align}
and so on.
On the other hand, we can 
exploit the Vi\`{e}te-Newton theorem \cite{wang2006solving} to 
build a polynomial $\mc P(x)$ with roots
$x=\alpha_{1,2,3,4}^{2}$ compatible with the constraints on $Q_{n\ge 2}$, see for instance (\ref{B.2}). 
In particular, 
the coefficients of the 
terms $1, x, x^{2}, x^{3}, x^{4}$ are fixed by $Q_{2,3,4,5}$. 
Instead, the values of $Q_{n\ge 6}$ must be compatible with the claim that 
$\mc P(x)$ has degree 4. These conditions impose algebraic constraints 
between the three quantities $Q_{1}$ , $\nu_{1}$, and $\beta$. 
For example, the condition that $\mc P(x)$ has no
$x^{5}$ term reads
\begin{align}
\la{B.3}
0&= -\frac{11 \beta ^{10}}{768}-\frac{125}{64} \beta ^7 \nu _1^2+\beta ^8 \left(\frac{5
   Q_1}{48}-\frac{55}{512}\right)+\beta ^6 \left(-\frac{55 Q_1^2}{192}+\frac{235
   Q_1}{384}-\frac{77}{384}\right)\notag \\
   &+\beta ^5 \left(\nu _1^2 \left(\frac{185
   Q_1}{32}-\frac{1415}{128}\right)-\frac{655 \nu _1^4}{128}\right)+\beta ^4
   \left(-\frac{75 \nu _1^4}{8}+\frac{35 Q_1^3}{96}-\frac{455 Q_1^2}{384}+\frac{187
   Q_1}{192}-\frac{77}{384}\right)\notag\\
   &+\beta ^3 \left(-\frac{105 \nu _1^6}{32}+\nu _1^4
   \left(\frac{25 Q_1}{4}-\frac{2125}{128}\right)+\nu _1^2 \left(-5 Q_1^2+\frac{515
   Q_1}{32}-\frac{1415}{128}\right)\right) \\
   &+\beta ^2 \left(-\frac{75 \nu _1^6}{16}+\nu
   _1^4 \left(\frac{45 Q_1}{8}-\frac{75}{8}\right)-\frac{5 Q_1^4}{24}+\frac{85
   Q_1^3}{96}-\frac{455 Q_1^2}{384}+\frac{235 Q_1}{384}-\frac{55}{512}\right)\notag \\
   &+\beta 
   \left(-\frac{45 \nu _1^8}{128}+\nu _1^6 \left(\frac{35
   Q_1}{32}-\frac{105}{32}\right)+\nu _1^4 \left(-\frac{25 Q_1^2}{16}+\frac{25
   Q_1}{4}-\frac{655}{128}\right)\notag \right. \\
   & \left. +\nu _1^2 \left(\frac{5 Q_1^3}{4}-5 Q_1^2+\frac{185
   Q_1}{32}-\frac{125}{64}\right)\right)+\frac{Q_1^5}{24}-\frac{5 Q_1^4}{24}+\frac{35
   Q_1^3}{96}-\frac{55 Q_1^2}{192}+\frac{5 Q_1}{48}-\frac{11}{768},\notag
\end{align}
with similar conditions from the vanishing of the $x^{6}$ and $x^{7}$ terms.
The solution of this reduced problem is affordable and for each 
solution $(Q_{1}, \nu_{1}, \beta)$ we obtain a specific polynomial $\mc P(x)$ whose roots 
can be identified with $\alpha_{i}^{2}$.

%
%
%
%
%
%
%
%
%
%
%
%
%
%
%
%
%
%
%
%
%
%
%

\newpage
\bibliography{N2-Biblio}
\bibliographystyle{JHEP}

\end{document}